\documentclass[aps,prd,twocolumn,groupedaddress,nofootinbib]{revtex4}
\usepackage{graphicx}
\usepackage{amssymb,amsfonts}
\begin{document}
\def\u{{\bf u}}
\def\k{{\bf k}}
\def\m{{\bf m}}
\def\ps{{\bf \cdot}}
\def\nb{\nabla}
\def\xp{x_p^{\rm eq}}
\def\ov{\over}
\def\l({\left(}
\def\r){\right)}
\def\la{\left[}
\def\ra{\right]}
\def\bc{\begin{center}}
\def\ec{\end{center}}
\def\be{\begin{equation}}
\def\ee{\end{equation}}
\def\bi{\begin{itemize}}
\def\ei{\end{itemize}}
\def\dt{\delta}
\def\Dt{\Delta}
\def\alf{\alpha}
\def\gam{\gamma}
\def\th{\vartheta}
\def\ph{\varphi}
\def\pt{\partial}
\def\eps{\varepsilon}
\def\omg{\Omega}
\def\df{\widehat{=}}
\def\mn{\mu\nu}
\def\dv{\rm{div}}
\def\d{\rm{d}}
\def\oa{\overrightarrow}
\def\vnb{\oa{\nabla}}


\title{Inertial modes in stratified rotating neutron stars : An evolutionary
description}

\author{L. Villain$^{1,2,3}$, S. Bonazzola$^1$ and P. Haensel$^2$}
\affiliation{\vspace*{0.2cm}$^1$Laboratoire de l'Univers et de ses Th\'eories
(UMR 8102 du C.N.R.S.), Observatoire de Paris -- \\
Section de Meudon, F-92195 Meudon Cedex, France\\
 $^2$Copernicus Astronomical Center (CAMK), Polish Academy of Sciences,
Bartycka 18, PL-00-716 Warszawa, Poland\\
 $^3$Departament d'Astronomia i Astrof\'{\i}sica, Universitat de Val\`encia,
E-46100 Burjassot, Spain\vspace*{0.2cm}}

\date{\today}

\begin{abstract}
 With (nonbarotropic) equations of state valid even when the neutron,
 proton and electron content of neutron star cores is not in beta
 equilibrium, we study inertial and composition gravity modes of
 relativistic rotating neutron stars. We solve the relativistic Euler
 equations in the time domain with a three dimensional numerical code
 based on spectral methods, in the slow rotation, relativistic Cowling
 and anelastic approximations. Principally, after a short description
 of the gravity modes due to smooth composition gradients, we focus
 our analysis on the question of how the inertial modes are affected
 by nonbarotropicity of the nuclear matter. In our study, the
 deviation with respect to barotropicity results from the frozen
 composition of nonsuperfluid matter composed of neutrons, protons
 and electrons, when beta equilibrium is broken by millisecond
 oscillations. We show that already for moderately fast rotating stars
 the increasing coupling between polar and axial modes makes those two
 cases less different than for very slowly rotating stars. In
 addition, as we directly solve the Euler equations, without coupling
 only a few number of spherical harmonics, we always find, for the
 models that we use, a discrete spectrum for the $l\,=\,m\,=\,2$
 inertial mode. Finally, we conclude that, for nonbarotropic stars, the
 frequency of this mode, which is our main focus, decreases in a
 non-negligible way, whereas the time dependence of the energy
 transfer between polar and axial modes is substantially different due
 to the existence of low-frequencies gravity modes.
\end{abstract}


\maketitle

\section{Introduction}  \label{sec:intro}

   Since the result of Andersson \cite{andersson98} and Friedman \&
Morsink \cite{friedman98}, numerous works have dealt with inertial
modes of rotating relativistic stars (see reviews by Friedman \&
Lockitch \cite{friedman01}, Andersson \cite{andersson03a} and Kokkotas
\& Ruoff \cite{kokkotas03}). In perfect fluids, these modes have
indeed been proven to satisfy the Chandrasekhar-Friedman-Schutz (CFS;
\cite{chandra70} and \cite{friedman78}) criterion for instability
whatever the angular velocity of the star is. Hence, due to their
coupling to the gravitational field, these oscillations make neutron
stars (NSs) very promising sources of gravitational waves (GW). Yet,
as NSs are complex relativistic objects, far from perfect fluids, the
astrophysical relevance of the instability of inertial modes in
rotating NSs is still an open issue. The final answer to this question
will depend on the many physical phenomena that take place within NSs:
their initial state at birth \cite{villain04}, existence of
differential rotation \cite{stergioulas04}, of a crust
(\cite{bildsten00}, \cite{lindblom00}, \cite{wu01}, \cite{yoshida01}),
of a huge magnetic field \cite{kinney03}, of superfluid states of
nucleons (\cite{yoshida03}, \cite{yoshida03b}, \cite{prix03}), of
exotic particles, {\it etc}. Understanding better that potential
instability is important for two main reasons. First, if it were
proven to be relevant, it could have a great impact on our idea about
the evolution of NSs and might explain the relatively slow angular
velocity of observed pulsars. Moreover, in order for the data
collected by GW detectors like VIRGO or LIGO to be useful, they have
to be compared with predicted signals, such as the signal that results
from the instability of inertial modes. Since theoretical predictions
depend on the models used for the inner structure of NSs, this
confrontation could possibly be as instructive for nuclear physics at
high density as was helioseismology for neutrino physics.\\

   In a previous article \cite{villain02} (referred to here as Paper
I), we have presented a spectral hydrodynamical code, which uses
spherical coordinates, with the aim of studying time evolution of
inertial modes in slowly rotating neutron stars. Having discussed the
motivation for the slow rotation approximation in this previous
article (it is still good even for the fastest known pulsars, see
Paper I), we keep working with it in the present paper. Thus, the next
step in our project was to improve the linear study of relativistic
inertial modes by trying to take into account in a more realistic way
the microphysical conditions that occur inside actual NSs. Among them,
we retained the stratification of the star. It was quite easy to
implement, and it is well-known that inertial modes change drastically
depending on whether the equation of state (EOS) that describes the NS
is barotropic or nonbarotropic\footnote{We remind that an equation of
state is said ``barotropic'' if the pressure $P$ depends only on one
variable, for instance the baryonic number density $n_b$.}, while the
status of purely axial modes of relativistic nonbarotropic NSs is
still under debate (\cite{friedman01},\cite{kokkotas03}). In addition,
it is easy to show that, for the most basic model of a star built with
the so-called ``$npe$'' matter, the assumption of a barotropic EOS,
even if easier to use, is wrong as soon as the star is cold enough and
the matter not in a stationary state. Hence, here we consider the time
evolution of inertial modes in slowly rotating relativistic NSs
composed of neutrons, protons and electrons, with an
``out-of-equilibrium'' EOS. It should be kept in mind that in the
following, we shall call ``nonbarotropic'' a peculiar type of
nonbarotropic EOSs: the situation of $npe$ matter with frozen
composition.\\

   The plan of the paper is as follows. In Section \ref{sec:bg}, we
start with an overview of the basic assumptions made in the study of
the structure of cold neutron stars, with emphasis on the beta
equilibrium between neutrons, protons and electrons, which constitute,
for not too massive NSs, the most important part for studies of their
oscillations: their core. We end this section with a description of
our calculation of stationary rotating NSs and of the EOS used to get
configurations to perturb in the following sections: the
Prakash-Ainsworth-Lattimer (PAL) EOS \cite{prakash1988}. In Section
\ref{sec:pertmicro}, a short summary of the microphysics of
``perturbed'' NSs composed of $npe$ matter is given. In this section,
we justify the assumption of frozen composition, which is then used in
Section \ref{sec:perthydro} that describes the basis of the linear
hydrodynamical study. The main feature of a nonbarotropic star is the
existence of a nondegenerate spectrum for its polar modes, even when
it is not rotating. As frozen composition results in a nonbarotropic
effective EOS, these so-called gravity modes ({\it g}-modes) are
studied for the PAL EOS in Section \ref{sec:gmod}. After which, in
Section \ref{sec:rotmod}, we describe inertial modes of rotating NSs
with the PAL EOS but without the (correct) hypothesis of frozen
composition, before ending with gravity and inertial modes in rotating
NSs taking into account the frozen composition. Finally, the results
are discussed in the Conclusion.

\section{Fully catalyzed rotating relativistic neutron stars} \label{sec:bg}

\subsection{Model of the neutron star matter} \label{sec:npe.model}

 The outer layers of cold NSs are known to play key roles in several
  physical phenomena, as the early and late cooling of pulsars
  \cite{yakovlev2001}, the existence of glitches (\cite{baym69},
  \cite{anderson75}, \cite{andersson03b}) and the spectra of X-ray
  bursts \cite{cottam02}. Yet, since more than $95\%$ of the mass of
  cold NSs is included in their core, outer layers are not relevant
  at the most basic levels of studies of both the structure of
  rotating relativistic stars and their oscillations. Thus, in the
  following, a NS is modeled by its core, and the outer layers are only
  discussed as time independent boundary conditions for the
  oscillations (see Section \ref{sec:gmod}).\\

  The minimal nontrivial description of a core of NS is a mixture of
  neutron, protons, and electrons at a temperature that is
  sufficiently low for the matter to be degenerate but sufficiently
  high for the nucleons not to be superfluid. Such conditions,
  which imply that thermal contributions to the EOS can be neglected,
  are satisfied for $10^9~K < T < 10^{10}~K$, at densities above one
  tenth of normal nuclear density $n_0\,\df\,0.16~{\rm fm}^{-3}$. Moreover,
  when a NS has cooled down enough for these assumptions to be valid,
  the matter has become transparent for neutrinos, which therefore do not
  contribute to the EOS either.\\

  Another physical constraint to take into account for the study of
  both unperturbed and perturbed stars is the electrical charge
  neutrality, which comes from the very short characteristic time
  scale of electromagnetic processes. In degenerate $npe$ matter, the
  so-called plasma frequency of electrons, whose inverse give the
  typical time for the plasma to maintain charge neutrality, is
\begin{equation}
\omega_{{\rm p}e}\,=\,\left(\frac{4 \pi e^2 n_e c^2}{\mu_e}\right)^{1/2}\,,
\end{equation}
 where $\mu_e$ is the chemical potential of electrons similar to
 $\mu_e\,\sim\,122\,(n_e/n_0)^{1/3}$ MeV (see below), with the
 electron density $n_e$. This leads to an electromagnetic relaxation
 time \mbox{$\tau_{{\rm p}e}\,=\,4\,\times 10^{-22}(n_e/n_0)^{1/3}~$s}, which
 implies that, for our purpose (phenomena with typical time scales of
 some ms), matter can always be considered as neutral, and there is
 equality between the electron and proton densities
 $n_e\,=\,n_p$. This relation also makes thermodynamical quantities
 functions of nucleon (baryon) density $n_b\,\df\,n_n\,+\,n_p$ and
 proton fraction $x_p\,\df\,n_p/n_b$ only. Furthermore, at baryon
 densities of the order of $n_0$, nucleons form a strongly interacting
 Fermi system and matter is very rich in neutrons, while we can
 replace nucleon masses (which are assumed to be equal) by the neutron
 mass $m_N\,=\,m_n=939.57$ MeV. If we neglect now the Coulomb
 interactions between charged constituents of matter and treat the
 electrons as a free ultrarelativistic Fermi gas, the energy per
 nucleon (excluding rest energy of nucleons) can be written
\begin{equation}
E(n_b,x_p)\,=\,E_{\rm N}(n_b,x_p)+E_e(n_b,x_p)~,
\label{e:E.Ne}
\end{equation}
where $E_{\rm N}$ is the nucleon contribution, and the electron
term $E_e$ is
\begin{equation}
E_e(n_b,x_p)\,=\,{3\over 4} b u^{1\over 3} x_p^{4\over 3}~, ~~~b\,\df\,\hbar\,
c\,(3\,\pi^2\,n_0)^{1/3}\,=\,331.4~{\rm MeV},
\label{e:Ee}
\end{equation}
with $u\,\df\, n_b/n_0$.\\ 

  Extensive many-body calculations with realistic two-nucleon and
three-nucleon forces show that in the whole range of $0\le x_p \le 1$,
the energy per nucleon is (to a very good approximation) quadratic in
the neutron excess parameter $(n_n-n_p)/n_b\,=\,1-2x_p$
(\cite{wiringa1988}, \cite{akmal1998}). At $x_p\simeq 0.5$,
characteristic value for terrestrial nuclear physics, this is just the
consequence of the charge symmetry of strong-interaction. At
$x_p\simeq 1$, this is a numerical fact. Therefore, the
$x_p$-dependence factors out, and $E_{\rm N}$ is given by the formula
\begin{equation}
E_{\rm N}(n_b,x_p)\,=\,W_0(n_b)\,+\,S(n_b)(1-2x_p)^2~.
\label{e:EN.W0.S}
\end{equation}
Here, $W_0(n_b)$ is the energy per nucleon of the symmetric nuclear
matter ($x_p={1\over 2}$), and $S(n_b)$ is the {\it symmetry energy}
at nucleon density $n_b$. Many possibilities exist for these two
functions, among them those that form the PAL EOS \cite{prakash1988}.
This EOS, which is retained in this study, will be described in more detail
with the presentation of the obtained background stars used for our
modes calculations (Section \ref{sec:bgm}). However, to be able to
calculate these unperturbed stars, another relation is needed to close
the system of equations. This relation comes from the hypothesis of
beta equilibrium between neutrons, protons and electrons.

\subsection{EOS in beta equilibrium} \label{sec:EOS.beta.eq}

 For matter transparent to neutrinos, the equilibrium with respect to
 the weak-interaction beta-processes
\begin{equation}
n\longrightarrow p + e +\overline{\nu}_e~,~~~
p+e \longrightarrow n +{\nu}_e~,
\label{e:beta.proc}
\end{equation}
implies a relation\footnote{Note that this relation remains the same
even when direct Urca processes, presented here, are not allowed due
to kinematical reasons, and have to be replaced by processes that
involve spectator nucleons, the modified Urca reactions.} between the chemical
potentials of the matter constituents, $\mu_j\,\df\,\partial \rho/{\partial n_j}$, with
\be
\rho(n_b,x_p)\,=\,n_b(m_N\,c^2\,+\,E)
\ee
the total energy density [energy per baryon $E$ that comes from
Eq.(\ref{e:E.Ne}) plus nucleon rest energies], and $j=n,p,e$. For
matter transparent to neutrinos, this relation reads
\begin{equation}
\mu_n\,=\,\mu_p+\mu_e~.
\label{e:beta.mu.eq}
\end{equation}

With our quadratic form of the $x_p$-dependence of $E_{\rm N}$,
Eq.(\ref{e:EN.W0.S}), the beta equilibrium condition involves only the
nuclear symmetry term. We have
\begin{equation}
\mu_n-\mu_p-\mu_e\,=\,4\,S\,(1-2x_p)\,-\,b\, u^{1\over 3} x_p^{1\over 3}~.
\label{e:beta.eq}
\end{equation}

 The resulting fraction of protons for matter in beta equilibrium,
 denoted by $\xp(n_b)$, is determined from
\begin{equation}
{{\xp}^{1\over 3}\over 1-2\,\xp}\,=\,
{4\,S\over b u^{1\over 3}}~,
\label{e:x.eq.n}
\end{equation}
or
\be \label{e:beteqy}
y^3\,+\,\alpha\,y\,-\,\frac{1}{2}\,=\,0,
\ee
where $y^3\,\df\,\xp$ and $\alpha(n_b)\,\df\,(b\,u^{1/3})/(8\,S)$.\\

 The last equation admits analytical solutions ({\it cf.} Cardan's method)
among which the only real one is
\be \label{e:xbeta}
x_p^{\rm eq}(n_b)\,\df\,\frac{1}{2}\,-\,\frac{\alpha(n_b)}{2^{2/3}}\,
\left[(\Gamma(n_b)+1)^{1/3}\,-\,(\Gamma(n_b)-1)^{1/3}\right]
\ee
with $\Gamma(n_b)\,\df\,\sqrt{1\,+\,\frac{16}{27}\,\alpha(n_b)^3}$.\\

 Then, the barotropic EOS in beta equilibrium, which is used for the
calculation of unperturbed stars, is obtained from 
\be \label{e:rho.bar.eos}
\rho_{\rm eq}(n_b)\,=\,\rho(n_b,\xp(n_b))\,,
\ee
with the total pressure
\be \label{e:p.bar.eos}
P_{\rm eq}(n_b)\,=\,n_b^2\,\left.\frac{\pt (\rho_{\rm eq}/n_b)}{\pt n_b}
\right|_{x_p}\,.
\ee

 Note that when the chemical potential of the electrons is above the
rest mass of muons, \mbox{$m_{\mu}\,\sim\,105.65$ MeV}, taking into
account the appearance of the latter via the equilibrium between the
processes
\be
e^-\,\to\,\mu^-\,+\,\nu_e\,+\,\bar{\nu}_{\mu}\hspace{0.3cm}
\textrm{and}\hspace{0.3cm}\mu^-\,\to\,e^-\,+\,\bar{\nu}_e\,+\,\nu_{\mu}\,,
\ee
would have implied higher proton fractions. But as a rule, the role
of muons is practically irrelevant, as far as the EOS is concerned,
and we shall always neglect them.

\subsection{Stationary rotating relativistic neutron stars} \label{sec:equa}

  The numerical solution of the Einstein's equations for a cold
stationary rotating relativistic star with a barotropic EOS, $P(\rho)$
[reached by Eq.(\ref{e:rho.bar.eos}) and (\ref{e:p.bar.eos})],
has now become a classical problem (see \cite{stergioulas03} for a
review) whose detail we will not enter into here. Let us just remind
the reader of the main assumptions:
\begin{description}
\item[-] there are two Killing vectors $\k$ (timelike at spacelike
infinity) for stationarity and $\m$ (with closed orbits, spacelike
everywhere except on the so-called rotation axis where it vanishes)
for axisymmetry;
\item[-] spacetime is asymptotically flat, the metric ${\bf g}$
``tends toward'' the flat Minkowski metric ${\bf \eta}$ at infinity;
\item[-] matter is a perfect fluid whose energy-momentum tensor
is\footnote{Except where otherwise noted we use natural units in which
\mbox{$c\,=\,{\cal G_N}\,=1$}.}
\be \label{e:perfflu}
{\bf T}\,\df\,(\rho\,+\,P) \u \otimes \u\,+\,P\,{\bf g}\,,
\ee
where $\u$ is the $4$-velocity of the fluid. Obviously, $P$ and $\rho$
are measured in the frame comoving with the fluid and thus are the
equilibrium functions obtained in the previous section;
\item[-] the velocity field corresponds to rigid axisymmetric
rotation and can be written
\be \label{e:dec_u}
\u\,=\,\lambda\l(\k\,+\,\omg\,\m\r)\,,
\ee
where $\omg$ is by definition the angular velocity of the star and
$\lambda$ a scalar field such as the norm of $\u$ is $-1$ (with our
convention for the signature).
\end{description}

 With these assumptions, the energy-momentum tensor satisfies the
condition of circularity \cite{carter69} and the {\it maximal slicing
quasi-isotropic} coordinates \cite{bonazzola93} can be adopted to
write the metric as
\begin{widetext}
\be \label{e:met_msqi}
{\d} s^2 \, \df \, g_{\mn}\, {\d} x^{\mu} {\d} x^{\nu}\,
\df \, - \l(N^2\,-\,N_{\ph}\,{N^{\ph}}\r) {\d t}^2 - 2\,N_{\ph}\, {\d t}
\,{\d\ph}\,+\,\frac{A^4}{B^2}\l({\d r}^2\,+\,r^2\,{\d\th}^2\,+\,B^4\,r^2\,
\sin^2 \th\,{\d\ph}^2\r)\,,
\ee
\end{widetext}
where we have introduced the notation of the $3+1$ formalism:
$N$ is the lapse and $N^{\ph}$ the third component of the shift
$3$-vector, with \mbox{$N_{\ph}\,\df\,g_{\ph i}\,N^i\,=\,A^4\,B^2\,
N^{\ph}r^2\,\sin^2 \th$} the covariant $\ph$-component of the
latter.\\

 Because of the existence of two Killing vectors all the functions only
 depend on ($r\,,\th$), and this choice of coordinates reduces the
 Einstein's equations to a system of elliptic equations that we solve with
the fully relativistic spectral code presented in Bonazzola
 {\it et al.} \cite{bonazzola93}. This code also provides
us with what will be the velocity field to perturb in the study of
oscillations (Section \ref{sec:perthydro}).
 Indeed the $\lambda$ scalar field, which normalizes the velocity, can
 be shown to satisfy
\be \label{e:lbd}
\lambda\,=\,-\left(|\k\ps\k\,+\,2\,\omg\,\k\ps\m\,+
\,\omg^2\,\m\ps\m|\right)^{-\frac{1}{2}}\,,
\ee
where ${\bf \cdot}$ denotes the scalar product associated with
the metric.

\subsection{Calculated background stars} \label{sec:bgm}

  In order to study inertial modes of NSs more realistic than simple
barotrops, we need an EOS for the nuclear matter that is also valid
when the fluid is not in beta equilibrium (\ref{e:beta.proc}). Indeed,
as explained in the Introduction and as calculations will show in
Section \ref{sec:pertmicro}, a lump of matter transported out of its
equilibrium position by an oscillation will not adjust its composition
to its new environment instantaneously. This will require some time,
which has to be taken into account in the dynamics since it allows new
modes (see Section \ref{sec:perthydro}) to exist.\\

  Among the numerous EOSs that describe nuclear matter in NSs, we
decided to use the equation proposed by Prakash {\it et al.}
\cite{prakash1988} which is among the few which are valid also out of
beta equilibrium. Moreover, this analytical EOS based partly on the
experimental properties of nuclear matter, and partly on the results
of many-body calculations of asymmetric nucleon matter, is in fact a
set of EOSs. Indeed, several possibilities are given for the choice of
the symmetry energy function $S(n_b)$, while the energy per nucleon of
the symmetric nuclear matter $W_0(n_b)$ depends on some
parameters. Thus, following Prakash {\it et al.}, we have
\begin{widetext}
\be \label{e:pal1}
W_0(n_b)\,\df\,\frac{3}{5}\,E_F^{(0)}\,u^{\frac{2}{3}}\,
+\,\frac{1}{2}\,A\,u\,+\,\frac{B\,u^{\sigma}}{1\,+\,B'\,u^{\sigma\,-\,1}}\,
+\,3\,\sum\limits_{i\,=1,2}\,C_i\,\l(\frac{\Lambda_i}{{p_F}^0}\r)^3\,
\l(\frac{p_F}{\Lambda_i}\,-\,\arctan\left[\frac{p_F}{\Lambda_i}\right]\r)\,,
\ee
\end{widetext}
where
\begin{description}
\item[-] $E_F^{(0)}$ is the free nucleon gas Fermi energy at
saturation density ($\sim 36$ MeV);
\item[-] the dimensionless variable $u$ is $u\,\df\,n_b\,/\,n_0$, where
\mbox{$n_0\,=\,0.16\,\rm{fm}^{-3}$} is the saturation density of symmetric
nuclear matter;
\item[-] the second and third terms, that depend on the constants $A,
B, B'$ and $\sigma$, reproduce the static interactions between
nucleons and are chosen in such a way that the EOS is causal;
\item[-] the last two terms, that depend on the momentum, mimic the
dynamical part of the strong interaction. In those terms, the
$\Lambda$ parameters are characteristic lengths associated to coupling
constants $C_i$. The first term (such as $C_1<0$ and
$\Lambda_1\,=\,1.5\,{p_F}^0$) is attractive at large distances, whereas
the second ($C_2\,>\,0$ with $\Lambda_2\,=\,3\,{p_F}^0$) reproduces
the repulsive behavior of the strong interaction at short scales.
\end{description}

 In order to separate the kinetic and potential contributions in
the symmetry energy, Prakash {\it et al.} \cite{prakash1988} write
\be
S(n_b)\,=\,\l(2^{\frac{2}{3}}\,-\,1\r)\frac{3}{5}\,E_F^{(0)}\,
\l(u^{\frac{2}{3}}\,-\,F(u)\r)\,+\,S_0\,F(u)\,,
\ee
where $F(u)$ is a function which describes the (poorly known) density
dependence of $S$. To get \mbox{$S(n_0)\,=\,S_0\,=\,30$ MeV},
one has $F(1)\,=\,1$.\\

 Playing with the form of the functions $W_0(n_b)$ and $S(n_b)$
enables one to change the values of the compression modulus and the
symmetry energy. However, this analytical EOS, that fits known results
around the saturation density, has the drawback that, at low
densities, it is no longer physical. Indeed, to be realistic at low
densities, any EOS has to take into account the appearance of
nuclei. Hence, for several choices of the key quantities (such as the
value of the compression modulus and the symmetry energy) within the
proposal of Prakash {\it et al.}, pathological features like negative
pressure or convective instability appear. Moreover, even for the EOSs
that do not present these awkward characteristics, the low density
part is unphysical and some prescription has to be adopted for the
study of modes.\\

 Among the functions and the set of parameters that Prakash {\it
et al.} propose, we retain $F(u)\,=\,\sqrt{u}$ and take values of
the parameters that give a compression modulus of either 180 MeV
(model A) or 240 MeV (model B). In addition, to prevent the troubles
linked to low densities, our choice is to {\it cut the star} below a
given threshold density. As discussed in Section \ref{sec:bg}, since
the mass of the crust is quite small, this cutoff does not really
affect either the metric or the main properties of the star. Moreover,
adding some boundary conditions on the surface described by the value
of the density chosen for the cutoff is an easy way to mimic the phase
transition that occurs between the outer-core and the
inner-crust. This point will be discussed in more detail in the
sections that deal with modes (Sections \ref{sec:perthydro},
\ref{sec:gmod}, \ref{sec:rotmod}) but it was noticed now to clarify
 Table \ref{tab:bg}. Indeed, in this table are summarized the
background configurations that we use. But in order to calibrate the
effect of the cutoff on the modes, we tried several values for the
density threshold. Hence, in Table \ref{tab:bg}, we show structural
properties of stars for the two EOSs that we keep with different central
densities and the values that we take for the cutoff density\footnote{{\it i.e.}
one half or almost two thirds of the saturation density $n_0$. This range is
consistent with realistic evaluations of the density at crust-core
interface (see \cite{douchin00} and \cite{haensel01a})}:
$0.08$ or $0.1$ baryon.fm$^{-3}$. As a first
validation of the cutoff procedure, we can check that for given EOS
and central density, the cutoffs imply differences in the
gravitational mass of some few percent in the worst case. Finally,
note also that the central (maximal) proton fraction is always below
$1/9$, which forbids direct Urca processes for $npe$ matter (see also
Section \ref{sec:EOSandNon-eq.proc}).


\begin{table*}
\centering
\caption[]{Properties of the NSs that are used as backgrounds for the
calculation of inertial and gravity modes. The entries in the table
are: type of EOS (name of model); central relativistic enthalpy (dimensionless); central baryonic
density (in fm$^{-3}$); central proton fraction; density at the surface; enthalpy at the surface;
gravitational mass (in solar mass units), circumferential equatorial
radius (in km) and compactness ${\cal C} = M_\mathrm{G} / R_{eq}$ (in dimensionless units).
Note that Model B is stiffer and allows a minimal gravitational mass which is larger
than $1.2$ M$_\odot$. This is the reason why we do not have stars
of model A and B with the same $1.2$ M$_\odot$ mass, whereas we have
stars of the same compactness A(26) and B(24), but also of the same
$1.6$ M$_\odot$ mass A(4) and B(3).}
\label{tab:bg}
\begin{tabular}{ c | ccc | ccc | ccc || ccc | ccc  }
  \hline
  \hline
 EOS  & \multicolumn{3}{c}{Model A(21)}  & \multicolumn{3}{c}{Model A(26)} & 
\multicolumn{3}{c}{Model A(4)} & \multicolumn{3}{c}{Model B(24)}
& \multicolumn{3}{c}{Model B(3)}  \\
  \hline
$h_\mathrm{c}$  &  \multicolumn{3}{c}{0.21}  &  \multicolumn{3}{c}{0.26}  &
\multicolumn{3}{c}{0.4}  &  \multicolumn{3}{c}{0.24}  &  \multicolumn{3}{c}{0.3}\\
$n_\mathrm{c}$ [fm$^{-3}$] &  \multicolumn{3}{c}{0.573}  &  \multicolumn{3}{c}{0.676}
&  \multicolumn{3}{c}{0.966} &  \multicolumn{3}{c}{0.540}  &  \multicolumn{3}{c}{0.635} \\
${x_p}_\mathrm{c}$ [\%]    &  \multicolumn{3}{c}{7.43}  &  \multicolumn{3}{c}{8.05}
&   \multicolumn{3}{c}{9.47} &  \multicolumn{3}{c}{7.22}  &  \multicolumn{3}{c}{7.81} \\
  \hline
$n_\mathrm{surf}$  [fm$^{-3}$] & 0 & 0.08 & 0.1   & 0 & 0.08 & 0.1   & 0 & 0.08 & 0.1   &
0 & 0.08 & 0.1    & 0 & 0.08 & 0.1\\
$h_\mathrm{surf}$[$10^{-2}$] & 0 & 2.0  & 2.3   & 0 & 2.0  & 2.3   & 0 & 2.0  & 2.3   &
0 & 1.9  & 2.2    & 0 & 1.9  & 2.2 \\
\hline
$M_\mathrm{G}$ [$M_{\odot}$] & 1.20 & 1.18 & 1.17 & 1.35 & 1.34 & 1.33 &
1.60 & 1.59 & 1.59 & 1.42 & 1.41 & 1.40 & 1.60 & 1.59 & 1.58 \\
$R_{eq}$ [km] & 11.3 & 10.4 & 10.3 & 11.1 & 10.4 & 10.3 & 10.2 & 10.0 &  9.9 & 11.8 & 11.1
& 11.0 & 11.6 & 11.0 & 10.9\\
${\cal C}$ [$10^{-1}$] & 1.57 & 1.67 & 1.68 & 1.79 & 1.91
& 1.91 & 2.33 & 2.36 & 2.38 & 1.77 & 1.88 & 1.89 & 2.03 & 2.13 & 2.14\\
\hline
  \end{tabular}
\end{table*}

\section{Perturbed configurations I: microphysics} \label{sec:pertmicro}

\subsection{$npe$ matter} \label{sec:npe_hydro}

  When trying to deal with modes of NSs in a not too unrealistic way,
{\it i.e.} when trying to go beyond the simple model of a single
barotropic fluid, the most basic model to consider is a NS composed of
neutrons, protons and electrons. Remember that we deal here only with
NSs sufficiently old to have become completely transparent to
neutrinos, but we neglect possible superfluidity of nucleons. Anyway,
even without neutrinos to complicate the game, these three fluids
could, {\it a priori}, be only partially coupled and then three
relativistic Euler equations (with coupling terms) could be
needed\footnote{Notice that in this case, the easiest way to reach the
equations of motion would not be to start with the usual conservation
of the energy-momentum tensor, but to use the Lagrangian formalism for
multifluids developed by Carter \cite{carter89}.}.\\

 But, as discussed in Section \ref{sec:bg}, due to the very short time
scale of the Coulomb interaction and due to its long range, it is
well-known that, for oscillations with periods typical of NSs modes
(some ms), protons and electrons are in fact interdependent with the
same velocities and densities: matter is locally electrically neutral
$x_p\,=\,x_e$. The first thing that we shall discuss now, is how the
strong interaction between nucleons plays a similar but not exactly
identical role. It makes the velocities of protons and neutrons
equal (without superfluidity), while not leveling out their
densities, as the latter depend on the weak-interaction (see Section
\ref{sec:EOS.beta.eq}). As the demonstration of this in the most
general case is quite tricky, we shall work here with a simpler
situation that provides an order of magnitude estimate sufficient for
our purpose.

\subsection{Damping of the relative $n-p$ flow}

 We assume that there is no external force acting differently on $n$
and $p$ ({\it i.e.}, we neglect the effects of the electromagnetic
field). Consider the relative motion of neutrons and protons along the
z-axis, with no baryon number flow, $v_n=-v_p=v$. Neglect all spatial
gradients, and assume, that at $t=0$ we have $v(0)=v_0$. By solving
the kinetic equations for neutron and proton distribution functions,
with a collision term due to the strong-interaction nucleon-nucleon
scattering, we will find that $v(t)=v(0)e^{-t/\tau_{np}}$, where
$\tau_{np}$ is the relaxation time after which the
relative $n-p$ flow is damped. As a nucleon fluid forms a strongly
interacting dense Fermi liquid, the $n-p$ scattering cross section to
be used in the kinetic equation is different from that in vacuum, and
the transport process refers to the quasiparticles of the Fermi liquid
theory (see, e.g., Baym {\it et al.} \cite{baym69}).\\

 For symmetric nuclear matter at normal nuclear density, one gets,
using the Fermi liquid theory, \mbox{$\tau_{np}\sim
10^{-18}\,T_9^{-2}$ s} (Haensel \cite{haensel1980}), where $T_9$ is
the temperature in $10^9$ K. For a highly asymmetric neutron star
matter, with $x_p\,=\,0.05$, simple estimate in which the vacuum $n-p$
scattering cross sections are used, were obtained by Baym {\it et al.}
\cite{baym69}, $\tau_{np}\sim 10^{-19}\,T_9^{-2}$ s. Hence, in view of
(microscopic) smallness of $\tau_{np}$ compared to any other time
scales relevant for our study, neutrons and protons can be assumed to
move together as one single nucleon fluid. Yet, as already mentioned,
the ratio between their densities is not governed by strong
interaction but by weak-interaction, which makes their relation
different from the relation between electrons and protons as we shall
see now.

\subsection{Nonequilibrium beta-processes} \label{sec:EOSandNon-eq.proc}

 Since the time for beta equilibrium (\ref{e:beta.proc}) to settle in
 depends on the production of thermal neutrinos, for the time being we
 shall consider an element of hot $npe$ matter at temperature $T$.
 This lump has an instantaneous nucleon density $n_b\,=\,n_p+n_n$,
 which due to its temporal evolution (that results from the matter
 flow) is out of beta equilibrium, so that
\begin{equation}
\dt\mu\,\df\, \mu_p + \mu_e - \mu_n \not= 0\,.
\label{e:delta.mu}
\end{equation}
Let us also characterize this deviation from beta
equilibrium by $\xi\,\df\,x_p\,-\,\xp(n_b)$.\\

 Each of the three fluids (neutrons, protons and electrons), is
separately in thermodynamic equilibrium, due to strong and
electromagnetic interactions. However, weak-interaction processes may
be too slow to establish beta equilibrium at [$n_b,T$]. We denote the
values of $n_n$ and $n_p$, which correspond to beta equilibrium at
[$n_b,T$], by $n_n^{\rm eq}$, $n_p^{\rm eq}$. The values $n_n^{\rm
eq}$, $n_p^{\rm eq}$ are the reference ones, for a given [$n_b,T$]
pair, and are compared with {\it actual} values $n_n$, $n_p$.\\

 A nonzero value of $\dt\mu$, Eq.(\ref{e:delta.mu}), implies
nonequilibrium reactions, which tend to decrease deviation from beta
equilibrium (in accordance with the Le Ch{\^a}telier principle). Let
us denote the rate of the change of $n_p$ at fixed $n_b$ by
$\Dt\Gamma(n_b,T,\dt\mu)$ (see Sect. 3.5 of Yakovlev {\it et al.}
\cite{yakovlev2001}). We then have
\begin{eqnarray}
&\Dt\Gamma(n_b,T,\dt\mu=0)=0~,\nonumber\\
&~~\Dt\Gamma(n_b,T,-\dt\mu)=-\Dt\Gamma(n_b,T,\dt\mu)~,
\label{e:Gamma}
\end{eqnarray}
while at fixed $n_b$, we can write
\begin{equation}
{\dot n_p}\,=\,-{\dot n_n}\,=\,{\dot\xi}\,n_b\,=\,\Dt\Gamma(n_b,T,\dt\mu)~,
\label{e:dot.n,p}
\end{equation}
where the dots denote time derivatives.\\

In the linear approximation, $\xi\ll \xp$, $\dt\mu$ implied by $\xi$
at a fixed $n_b$ is
\begin{equation}
\left(\dt\mu\right)_{n_b}=
\left({\partial \dt\mu\over \partial x}\right)_{n_b}\xi=
{4\over 3}S\left(4 +{1\over x}\right)\xi~,
\label{e:delta.mu.xi}
\end{equation}
where all coefficients are calculated at $n_b,\xp$.\\

In what follows, we assume in addition that $\dt\mu/(\pi kT)\ll 1$,
which is equivalent to \mbox{$\dt\mu/{\rm MeV}\ll 0.27~T_9$}. This means that
matter deviates only weakly from the beta equilibrium, an
approximation which is valid for small amplitude of neutron star
pulsation. Moreover, we remind the reader that nucleons are supposed
to be normal (no superfluidity, valid for $T_9>1$) and we will
consider here only cases with $x_p<1/9$, which is in perfect
agreement with our background models (see Table \ref{tab:bg}), so that
only modified Urca processes are allowed (see Lattimer {\it et al.}
\cite{lattimer1991}). Notice that the case of direct Urca processes
would anyway lead to a similar conclusion, just changing
quantitatively the result. Under the previous assumptions, the formula
for $\Dt \Gamma$ reads, keeping only terms linear in $\dt\mu$,
\begin{equation}
\Dt\Gamma(n_b,T,\dt\mu)=3.9\times 10^{28}~
\left({n_p\over n_0}\right)^{1/3}~T_9^6~
{\dt\mu\over {\rm 1~MeV}}
~~{\rm cm^{-3}~s^{-1}}~,
\label{e:Gamma.mu}
\end{equation}
where normal nuclear density $n_0=0.16~{\rm fm^{-3}}$.\\

In the relaxation time approximation, we can rewrite the formula
for $\Dt\Gamma$ as
\begin{equation}
\Dt\Gamma= -{{n_p-n_p^{\rm eq}}\over \tau_\beta}~.
\label{e:dot.n_n}
\end{equation}

At fixed $n_b$, relaxation from a state initially off equilibrium
by $\dt n_p^0\equiv n_p^0 - n_p^{\rm eq}$,
towards beta equilibrium, proceeds according to
\begin{equation}
n_p(t)\,=\,n_p^{\rm eq} + \dt n_p^0{\rm e}^{-t/\tau_\beta}~.
\label{e:relax}
\end{equation}

Using the relation between $\dt\mu$ and $\dt n_p$, one shows
that the beta--relaxation time is given by
\begin{equation}
\tau_\beta=5.1\times 10^5~(T_9)^{-6}
\left( {n_b\over n_0}\cdot {x_p\over 0.01}\right)^{1/3}~{\rm s}~.
\label{e:tau_beta}
\end{equation}

Under the prevailing conditions, this time scale is so long compared
to other typical time scales that it is a very good assumption to
consider that every lump of matter keeps a frozen composition (value
of $x_p$) in perturbed configurations\footnote{If hyperons were
included, they would also move together with the bulk matter due to
the strong interaction, but concerning the respective ratios, the
situation would be a little different due to nonleptonic strangeness
violating reactions. These weak-interaction reactions without leptons
like
\begin{equation}
 {\rm   N} \Lambda  \rightleftharpoons  {\rm N n}, \quad
   {\rm nn \rightleftharpoons  p} \Sigma^-\,,
\end{equation}
where N is a nucleon, have relaxation times of the order of the
millisecond for $T\,\sim\,10^9$ K
(\cite{haensel02},\cite{haensel02b}). Thus the composition could be no
longer frozen for millisecond oscillations.}. In Section
\ref{sec:perthydro} we shall see how, from the hydrodynamical point of
view, this condition can be taken into account and leads to a
nonbarotropic equation of state. Before that, we shall briefly
discuss for $npe$ matter viscosity coefficients and viscous damping
times of oscillations.

\subsection{Viscosity of $npe$ matter} \label{sec:visco}

 As we shall see later (Sections \ref{sec:gmod} and \ref{sec:rotmod}),
we deal in this study only with rotational and gravity $m\,=\,2$
modes, with the idea of focusing on the gravitational driven
instability of the {\it r-}modes. For them, it is well-known
\cite{kokkotas03} that shear viscosity kills the CFS instability (see
Section \ref{sec:rotmod}) for low temperatures, whereas bulk-viscosity
does it for high temperatures. On the other hand, gravity modes of
cold NS star are not directly relevant for gravitational waves
emission due to their very weak coupling with the gravitational
field. Hence our goal here is just to ``verify'' that to use the
equations of motion for perfect fluids is quite reasonable for the
durations of our simulations (of the order of 1 second) during which
viscosity would not really have time to damp the modes.\\

The dissipation connected with viscosity of the $npe$ matter is
characterized by two density and temperature dependent parameters,
shear viscosity $\eta$ and bulk-viscosity $\zeta$. Shear viscosity
results from the momentum transfer in the scattering processes between
the $npe$ matter constituents. It is mostly determined by
large-momentum transfer collisions, and therefore the dominant
contribution to $\eta$ is that of nonsuperfluid neutrons. An approximate
analytic expression describing results obtained by Flowers \& Itoh \cite{flowers1979} is
\begin{equation} \label{eq:eta}
\eta\simeq \eta_n\simeq 1.6\times 10^{16}(\rho_{14})^2~(T_9)^{-2}~{\rm
g~cm^{-1}~s^{-1}}~, 
\end{equation}
where $\rho_{14}\,\df\,\rho/10^{14}~{\rm g~cm^{-3}}$.\\

On the other hand, bulk-viscosity results from the nonequilibrium beta-processes
previously discussed. But remember that for the notion of bulk-viscosity to be
valid, the deviations from the beta equilibrium measured by
$\delta\mu$ should be much smaller than $k\,T$. In such conditions, by considering periodic
pulsations of local pressure in the $npe$ matter, one can obtain an expression for
the mean heat deposition due to the nonequilibrium processes. When the relation between
Fermi momenta
\be 
p_{{\rm F}n}>p_{{\rm F}p}+p_{{\rm F}e}\,
\ee
is fulfilled, which in the case of $npe$ matter corresponds to $x_p<1/9$, the direct
Urca processes are prohibited. As already mentioned above with the
description of our background stars, this is always the case in this
study (see Table \ref{tab:bg}). Hence here only modified Urca (mUrca)
processes are allowed, with their neutron branch
\begin{eqnarray} \label{e:mUrca.n}
n+n\longrightarrow n+p+e+\bar{\nu}_e~~, \cr
n+p+e\longrightarrow n+n+\nu_e~,
\end{eqnarray}
and their proton branch
\begin{eqnarray} \label{e:mUrca.p}
p+n\longrightarrow p+p+e+\bar{\nu}_e~~, \cr
p+p+e\longrightarrow p+n+\nu_e~.
\end{eqnarray}

 The formulae obtained by Haensel {\it et al.}
(see \cite{haensel01b},\cite{haensel00} for more detail) and applied
to the $npe$ matter give for the resulting bulk-viscosity
\begin{equation}
\zeta^{\rm mUrca}_n\,\sim\,10^{20}
 \omega_4^{-2} (T_9)^6 ~{\rm g~cm^{-1}~s^{-1}}~.
\end{equation}

Here, $\omega_4$ is the angular frequency of pulsations measured in the units of
$10^4~{\rm s^{-1}}$. Even if we shall not need it here, let us mention that the
magnitude of the bulk-viscosity increases by many orders if the
direct Urca (dUrca) processes are allowed. These processes lead to
\begin{equation} \label{eq:zeta}
\zeta^{\rm dUrca}_n\,\sim\,10^{25}
\omega_4^{-2}
(T_9)^4 \Theta_{npe}~{\rm g~cm^{-1}~s^{-1}}~,
\end{equation}
where the threshold factor $\Theta_{npe}$ vanishes if $p_{{\rm
F}n}>p_{{\rm F}p}+p_{{\rm F}e}$.\\

 With Eq.(\ref{eq:eta}), we obtain the time scale
for the shear viscosity \cite{cutler87}
\begin{equation}
\tau_\eta= \alpha^{\eta}_{\rm mode}
{\rho R^2\over \eta}\simeq \alpha^{\eta}_{\rm mode}
\;{R_6^2\,T_9^2\over \rho_{15}}\times 10^9~{\rm s}~,
\label{eq:tau.eta}
\end{equation}
where $R=R_6\times 10^6~{\rm cm}$ is the core radius,
$\rho=\rho_{15}\times 10^{15}~{\rm g~cm^{-3}}$ the ``mean'' core
density, and $\alpha^{\eta}_{\rm mode}$ a numerical
coefficient that depends on the pulsational mode.\\

 Eq.(\ref{eq:zeta}) give a very rough estimate of the bulk-viscosity damping
time scale due to mUrca bulk-viscosity:
\begin{equation}
\tau^{\rm mU}_\zeta\,\sim\,\alpha^{\zeta,{\rm mU}}_{\rm mode}
{\rho R^2\over \zeta^{\rm mU}}
\simeq \alpha^{\zeta,{\rm mU}}_{\rm mode}
\;{R_6^2\over  T_9^6}\times 10^8~{\rm s}~.
\label{eq:tau.zeta}
\end{equation}

We see with these formulae that around $10^9$ K, none of the
viscosities damps the modes very quickly\footnote{the expected value
of $\alpha^{\zeta,{\rm mU}}$ for the {\it r-}mode is significantly
greater than one.}. More precisely, for $T_9\,=\,5$ the damping time
scale is of the order of hours. Hence for an evolution lasting a few
seconds the assumption of perfect fluid is not so bad.\\

 As a final remark, let us mention that, in a star of radius $R$, the
presence of a core of radius $R_D$ where the direct Urca process is
allowed would imply a damping time scale given by
\begin{equation}
\tau^{\rm dU}_\zeta= \alpha^{\zeta,{\rm dU}}_{\rm mode}
{\rho R^2\over \zeta^{\rm dU}}
\simeq \alpha^{\zeta,{\rm dUrca}}_{\rm mode}
\;{R_{6}^2\over  T_9^4} \times 10^3~{\rm s}~,
\label{eq:tau.damp}
\end{equation}
with, for the {\it r}-modes, $\alpha^{\zeta,{\rm dU}}$ much greater
than one (see \cite{zdunik96} for the case of the CFS
instability). Additional increase of $\alpha^{\zeta,{\rm dUrca}}_{\rm
mode}$ results from the fact that the sphere in which energy is
dissipated has $R_D$ for radius, that is only a fraction of the
stellar radius $R$ in which the mode propagate. However, the direct
Urca process would not be for the modes as dramatic as the presence of
hyperons and at $T_9\,=\,5$ we would still expect damping on a time
scale of minutes.

\section{Perturbed configurations II: hydrodynamics} \label{sec:perthydro}

\subsection{Assumptions for the linear hydrodynamical study}

   Doing time evolutions in general relativity is much more complicated
 than the solution of Einstein's equations for stationary and
 axisymmetric configurations. Hence, it is natural (at least in a first
 study) to begin with some approximations in order to simplify the problem.
Thus, for the rest of this work, we assume that
\begin{description}
\item[-] the perturbed star is ``not too far'' (see \cite{friedman78}
for a proper definition of this) from the unperturbed one. It is then
legitimate to linearize all equations (that govern both the material
fields and the gravitational field) with respect to a parameter that
indexes a continuous family of time dependent solutions and whose
value is $0$ for the unperturbed configuration. Obviously, this
approximation means that in the following study of modes, the zero
order terms are input that come from the unperturbed calculation
presented in Section \ref{sec:bgm};
\item[-] the Eulerian perturbations of the gravitational field ({\it
i.e.} of the metric) are quite ``small'' and not fundamental for the
physics. We then neglect them and write
\be
\dt g_{\mn} \equiv 0\,,
\ee
where $\dt g_{\mn}$ are the Eulerian perturbations of the metric
$g_{\mn}$. This is the so-called relativistic Cowling \cite{cowling41}
approximation, introduced by Mc Dermott {\it et al.}
\cite{mcdermott83} in the study of oscillations of warm neutron
stars. As discussed by several authors, this approximation has
several drawbacks in the relativistic case (\cite{stergioulas04},\cite{finn88}),
but it was also shown that it does not change dramatically the main feature of
inertial modes (see \cite{kokkotas03} for a review). In addition, implementing the
evolution of some metric perturbations (see Section \ref{sec:rotmod})
in our code would require a non-negligible amount of work since
boundary conditions have to be treated carefully with spectral methods
\cite{novak04};
\item[-] the NS is slowly rotating. Only terms linear in the
angular velocity $\omg$ are kept in the equations, and every zero
order term is assumed to only depend on the radial coordinate. This
assumption was discussed in Paper I, and we shall stress here just one
point : its use makes it sufficient to work with Hartle's equations
\cite{hartle67} and not with the full Einstein's equations for the
unperturbed calculation. Yet, as we plan to get rid of it in the
future and since we have a fully relativistic code, it was easier for
us to use this code to get the zero order terms. Moreover, we verified
in this way the relevance of the slow rotation approximation. For more
detail, see Paper I;
\item[-] the spatial part of the metric is conformally flat
(Isenberg\--Wilson\--Mathews approximation, \cite{isenberg78},
\cite{isenberg80} and \cite{wilson89}), {\it i.e.} the metric
(\ref{e:met_msqi}) becomes
\be
{\d s}^2 \, = \, - \l(N^2\,-\,N_{\ph}\,{N^{\ph}}\r) {\d t}^2
- 2\,N_{\ph}\, {\d t}\,{\d\ph}\, + \, h_{ij}\, {\d} x^i \, {\d} x^j\,
\ee
with $i,j\,\in\,(1,2,3)$ and
\be
{\bf h}\,\equiv\,a\,{\bf \eta}\,,
\ee
where $a$ is the conformal factor and {\bf $\eta$} the flat Euclidean
$3$-metric. The conformal approximation was shown to be very good
for (even fast) rotating isolated NSs by Cook {\it et al.}
\cite{cook96}, and once again using the fully relativistic code we
could verify this before making the assumption in the linear
part of the work. Moreover, as we already use the Cowling
approximation, we do not really care to ``kill again'' the GW content
of spacetime, which can at a first level be obtained with some
``post-Newtonian multipolar scheme'' (see for instance
\cite{blanchet02}).
\end{description}

 With all these assumptions, the final frozen metric used in the first
order calculations is written
\begin{widetext}
\be \label{e:metric}
{\d s}^2\,\df\,\,- \l(N^2\,-\,a^2\,r^2\,{\sin[\th]}^2\,{N^{\ph}}^2\r) {\d t}^2
\,-\,2\, a^2\,r^2\,{\sin[\th]}^2 \,N^{\ph} {\d t}\,{\d\ph}\,+\,a^2\,{\d l}^2\,,
\ee
\end{widetext}
where any function $N, N^{\ph}$ and $a$ only depends on the radial
coordinate $x^1\,\df\,r$, while ${\d l}^2$ is the length interval in
the flat $3$-space. With this metric, the linearized equations of
motion can be easily written in a way similar to the Newtonian
equations.

\subsection{Equations of motion} \label{sec:eom}

 For reasons described in Section \ref{sec:pertmicro}, we
assume that there is only one fluid whose relativistic
equations of motion are obtained by the usual conservation of the
energy-momentum tensor of a perfect fluid
\be \label{e:cons_energ}
{\bf \nabla}\,\ps\,{\bf T}\,=\,0\,,
\ee
with
\be 
{\bf T}\,\df\,(\rho\,+\,P) \u \otimes \u\,+\,P\,{\bf g}\,,
\ee
where ${\bf \nabla}$ is the covariant derivative associated with the
Levi-Civita affine connection for the given metric (\ref{e:metric}).\\

 For the isentropic motion of a perfect fluid with a given EOS, it is
well-known that, due to thermodynamics, the system of equations formed
by the baryonic number conservation and the energy-momentum
conservation (\ref{e:cons_energ}) is degenerate. Hence, we shall work
with the baryonic number conservation,
\be \label{e:bar_cons}
{\bf \nabla}\,\ps\,\l(n_b\,\u\r)\,=\,0\,,
\ee
where $n_b$ is the baryonic density measured in the fluid's local rest
frame, plus the projections on the $3$-space of the relativistic
Euler equations (EE) obtained from Eq.(\ref{e:cons_energ}). More precisely, the useful equations
are those reached after linearizing close to a solution that describes the rigidly rotating
star, Eqs. (\ref{e:dec_u},\ref{e:lbd}).\\

  As was briefly described in Paper I, in order to get equations as
similar as possible to the Newtonian EE, we use the well-known results about
rigid rotation of relativistic stars, Eqs. (\ref{e:dec_u},\ref{e:lbd}),
and write the perturbed $4$-velocity as
\begin{widetext}
\be \label{e:4v_tot}
u^{\mu}[t,r,\th,\phi] \, = \frac{1}{N[r]}
\left| \begin{array}{ll}
1+\dt U^0[t,r,\th,\phi] & \\
{\l(N[r]\,/\,a[r]\r)}^2\,\dt U^r[t,r,\th,\phi] & \\
{\l(N[r]\,/\,a[r]\r)}^2 \, \dt U^{\th}[t,r,\th,\phi]\,/\,r& \\
\omg\,+\,{\l(N[r]\,/\,a[r]\r)}^2\,\dt U^{\ph}[t,r,\th,\phi]\,/\,(r \sin[\th])&
\end{array} \right.\,.
\ee
\end{widetext}

  Here, $\dt U^0, \dt U^r, \dt U^{\th}$ and $\dt U^{\ph}$ are first
order quantities added to the unperturbed $4$-velocity that describes
rigid rotation, with $\dt U^0$ that is not a dynamical variable but
determined according to the constraint that $\u$ is a $4$-velocity :
\mbox{$\u\,\ps\,\u\,=\,-1$} for the chosen signature of the metric.
Furthermore, $\dt U^r, \dt U^{\th}$ and $\dt U^{\ph}$ are not
contravariant components of a $4$-vector but convenient variables used
in our calculations, which are nevertheless the components of
$\oa{W}$, a $3$-vector, in the orthonormal basis associated with the
spherical system of coordinates for the flat $3$-space. With all
this, we obtain
\begin{widetext}
\begin{eqnarray} \label{e:eul_rel}
& \l(\pt_t\,+\,\omg\,\pt_{\phi}\r)\dt U^r\,+\,\dt \left[\pt_r P\,/\,f\right]\,
-\,2\,\dt U^{\ph}\,\sin[\th]\l(\l(\omg\,-\,N^{\ph}\r)
\l(r\,a'\,/\,a\,+\,1\,-\,r\,N'\,/\,N\r)\,-
\,r\,N^{\ph'}\,/\,2\r)\,=\,0\,,\nonumber
\end{eqnarray}
\begin{eqnarray}
&\l(\pt_t\,+\,\omg\,\pt_{\phi}\r)\dt U^{\th}\,+\,
\dt \left[\pt_{\th} P\,/\,(r\,f)\right]
-\,2\,\dt U^{\ph}\,\cos[\th]\l(\omg\,-\,N^{\ph}\r)\,=\,0\,,
\end{eqnarray}
\begin{eqnarray}
&\l(\pt_t\,+\,\omg\,\pt_{\phi}\r)\dt U^{\ph}\,+\,
\dt \left[\pt_{\ph} P\,/\,(f\,r\,\sin[\th])\right]\,+\,2\,\dt U^r\,\sin[\th]
\l(\l(\omg\,-\,N^{\ph}\r)\l(r\,a'\,/\,a\,+
\,1\,-\,r\,N'\,/\,N\r)\,-\,r\,N^{\ph'}\,/\,2\r)\,\nonumber\\
& +\,2\,\dt U^{\th}\,\cos[\th]\l(\omg\,-\,N^{\ph}\r)\,=\,0\,,\nonumber
\end{eqnarray}
\end{widetext}
where the $'$ indicates the derivation with respect to the radial
coordinate, $\dt$ is the operator for the Eulerian perturbation and
$f\,\df\,\rho\,+\,P$.\\

 To this system of equations, we add the linearized baryon number
conservation that, in the slow rotation approximation, can be
written as
\be \label{e:lin_bar_cons}
\l(\pt_t\,+\,\omg\,\pt_{\ph}\r)\,\dt \tilde{n}\,
+\,\frac{N^2}{a^2}\,
\textrm{div}\l(\tilde{n}\,\oa{W}\r)\,=\,0\,,
\ee
where we define $\tilde{n}\,\df\,n_b\,{N[r]}^2\,a[r]$ and introduce
the already mentioned $3$-vector $\oa{W}$, using also the usual
$3$-dimensional spatial divergence operator.\\

 Hence, with the introduction of the relativistic anelastic
approximation defined in Paper I (see also this paper for a discussion
of the motivations for this approximation), we finally get the
Newtonian-like equation
\be \label{e:ane_cons}
\textrm{div}\l(\tilde{n}\,\oa{W}\r)\,=\,0\,,
\ee
which is the equation used in the current version of the code. The
only missing element to close the system of equations is now the way
thermodynamical quantities (mainly the pressure) are perturbed, {\it
i.e.} we need to give a prescription to calculate in system (\ref{e:eul_rel})
the term
\be \label{e:pert_enth}
\dt \left[\frac{1}{f}\,\vnb P\right]\,.
\ee

 This prescription will come directly from the previous analysis and
 from the fact that matter keeps a frozen composition when it is perturbed.

\subsection{Motion of perturbed $npe$ matter} \label{sec:motnpe}

 Let us first summarize the conclusions that come from the
 microphysical study overviewed in Section \ref{sec:pertmicro}:
\begin{description}
\item[-] the unperturbed star is described by an effective barotropic
EOS with the $npe$ matter in beta equilibrium;
\item[-] the perturbed (oscillating) star can not be described by the
same EOS since matter is no longer in beta equilibrium;
\item[-] the easiest but most general EOS for perturbed $npe$ matter at low temperature
depends on two parameters: the baryonic density $n_b$ and the proton fraction $x_p$;
\item[-] for this nonbarotropic EOS and in the situation retained
for the present study, the physical conditions in which the
perturbations occur are that a given piece of $npe$ matter will move
with a frozen composition that corresponds to its composition in the
unperturbed situation.
\end{description}

 This last point needs now to be translated into an equation that will
 replace the beta equilibrium relation in the case of perturbed
 matter. But before we make it more explicit, we shall remind what
 turns out to be the situation when instantaneous beta equilibrium (or
 barotropicity) is assumed. This situation was supposed to occur in
 Paper I, and is still useful here since we have tried to modify our
algorithms as little as possible.\\

  In this case, thermodynamics tells us that both the pressure $P$ and
the total energy density $\rho$ are functions only of the baryonic
density $n_b$. Hence, the term (\ref{e:pert_enth}) can be written as
the Eulerian perturbation of an exact derivative (or the gradient of a
scalar field), or equivalently as the gradient of the perturbation of
a field that comes from an exact derivative, the relativistic
enthalpy
\be \label{e:rel_enth}
\dt \left[\frac{1}{f(n_b)}\,\vnb P(n_b)\right]\,\df\,\dt \vnb H(n_b)\,
\equiv\,\vnb \dt H(n_b)\,.
\ee

 Using this new function, one can write the EE in a condensed and
well-known form, which makes the system self-consistent and which
was described in Paper I.\\

 When we can no longer use the same EOS for the perturbed star
and the background star, the previous list of conclusions are
translated into equations in the following ways. First, as already
mentioned, they imply that we need an EOS for the perturbed fluid that
relates the perturbation of the pressure $P$ and energy density $\rho$
with (for instance) the perturbation of the baryonic density $n_b$ and
the perturbation of the proton fraction $x_p$
\be \label{e:pert_pr_nonbar}
\dt P(n_b,x_p)\,=\,\left.\frac{\pt P}{\pt n_b}\right|_{x_p}\,\dt n_b\,
+\left.\frac{\pt P}{\pt x_p}\right|_{n_b}\,\dt x_p\,,
\ee
and no longer only with the first one\footnote{From the relativistic
point of view, it is crucial to verify that the employed EOS dependent
on these parameters is covariant. It can be shown to be trivial if we
write the numbers $n_b$ and $x_p$ as Lorentz scalars defined with
the baryon, neutron and proton $4$-currents. It gives
\be
n_i\,\df\,\sqrt{-\,{{{\bf n_i}}\,\ps\,{{\bf n_i}}}}\,,
\ee
where $i$ is either $b$, $n$ or $p$. Moreover, with
the relations between $4$-currents and $4$-velocities
\be
{\bf n_i}\,=\,n_i\,{\bf u_i}\,,
\ee
it is easy to see that assuming each of the velocity fields to
be equal implies that the most general EOS can only
depend on two Lorentz scalars, $n_b$ and $x_p$ for instance.}
\be   \label{e:pert_pr_bar}
\dt P_{\beta}(n_b)\,\df\,\left.\frac{{\d} P}{{\d} n_b}\right|_{x_p\,\equiv\,
\xp(n_b)}\,\dt n_b\,\df\,\frac{P}{n_b}\,\gamma_{\beta}\,\dt n_b\,,
\ee
as it was when instantaneous beta equilibrium and a barotropic EOS
were assumed. Note that here we have introduced the adiabatic index
for the matter in beta equilibrium, $\gamma_{\beta}$. This index will be
used in the following and is {\it a priori} not a constant (as it is
in the polytropic case). Moreover, the function $\xp(n_b)$, which
links the proton fraction with the baryonic density for $npe$ matter
in beta equilibrium, is of course the result (\ref{e:xbeta}) obtained
in Section \ref{sec:bg}.\\

 One of the main implications of an EOS dependent on two parameters is
that it makes it impossible to define the relativistic enthalpy by the
relation (\ref{e:rel_enth}), since the fraction is no longer an exact
derivative. Hence, the term (\ref{e:pert_enth}) can no longer be
regarded as a function of a single variable and the system has to be
completed by an additional equation to be closed, an equation that
describes the fact that the matter has a frozen composition.\\

 This prescription means that the Lagrangian perturbation of the
proton fraction is $0$ (as far as the EE are concerned) or
(equivalently) that its derivative along the velocity field
vanishes. Addressing with $\Dt$ the Lagrangian perturbation of any
quantity, we then have for the motions that we are dealing with
\be
\Dt P\,=\,\frac{P}{n_b}\,\gamma_F\,\Dt n_b\,,
\ee
where
\be
\gamma_F\,\,\df\,\left.\frac{\pt \ln[P]}{\pt \ln[n_b]} \right|_{x_p}\,
\ee
is the adiabatic index for a frozen composition. Furthermore, the formal
relation, which links Lagrangian and Eulerian perturbations by the
Lie derivative with respect to the Lagrangian deplacement ${\bf \xi}$,
\be
\Dt - \dt \,\df\,\pounds_{{\bf \xi}}\,,
\ee
enables us to arrive to the final expression
\be \label{e:devp}
\dt P\,=\,\dt P_{\beta}\,+\,\frac{P}{n_b}\,\Dt n_b\,
\l(\gamma_F\,-\gamma_\beta\r)\,
\ee
where the first part of the perturbation is what we would get if beta
equilibrium was instantaneous. A similar expression can be found for
the perturbation of $f\,\df\,\rho\,+\,P$ with some ``$\zeta$''
coefficients instead of the $\gamma$. Nevertheless, the relevant quantity
at the linear order for matter close to beta equilibrium is
\be
\zeta_F\,-\zeta_\beta\,=\,\frac{P}{f}\,\l(\gamma_F\,-\gamma_\beta\r)\,.
\ee

 The expression (\ref{e:devp}) and its equivalent for $f$ are those we
use in the EE in order to write the term (\ref{e:pert_enth}) as the
perturbation of the gradient of the enthalpy plus a contribution that
exists only for nonbarotropic stars. In this way, we do not need to
change our algorithms (see Paper I for more detail) for the solution
of the EE and the anelastic equation (\ref{e:ane_cons}), whereas we
only need to add, as a source term, the second part of these
expressions. Moreover, written like this, this source term depends only on
the Lagrangian perturbation of the baryonic density $\Dt n_b$ (as far
as first order terms are concerned), since we have
\begin{eqnarray}
\label{e:source}
\dt \left[\frac{1}{f}\,\vnb P\right]\,=&\,\vnb \dt H\,
+\,\frac{P}{f}\,\vnb \l(\frac{\Dt n_b}{n_b}\,\l(\gamma_F\,-\gamma_\beta\r)\r)\,\nonumber\\
&+\,\vnb H\,\frac{\Dt n_b}{n_b}\,\l(\gamma_F\,-\gamma_\beta\r)\,\l(1\,-\,\frac{P}{f}\r)\,,
\end{eqnarray}
in which $\Dt n_b$ is obtained by solving the equation
\be \label{e:derlie_xp}
\pounds_{\u}\,x_p\,=\,0\,,
\ee
where $\u$ is the $4$-velocity. Indeed, when we linearize this equation
and use the anelastic approximation, we get after some algebra the
Newtonian-like advection equation for $\Dt n_b$,
\be
\l(\pt_t\,+\,\omg\,\pt_\ph\r)\,\Dt n_b\,=
\,\frac{N^2}{a^2}\,\oa{W}\,\cdot\,\vnb n_b\,,
\ee
where the usual Newtonian notations were used. This new equation
enables the existence of modes related to the gradient of composition
and whose restoring force is gravitation. These {\it g}-modes for cold
but heterogeneous NSs were first discussed by Reisenegger \& Goldreich
\cite{reisenegger92} and will be the subject of the next section.

\section{Gravity modes} \label{sec:gmod}

\subsection{General features of composition gravity modes}

  Before dealing with the more complex situation of the modes of a
stratified rotating relativistic star, it is worth to recall the
situation for the modes of a nonrotating star, in order to better
recognize in the following their scions. With the already mentioned
assumptions, it is quite simple to describe. We consider only the
$npe$ core of NSs, without superfluidity, without magnetic field and
without trying to follow modes of the crust. Hence due to the
spherical symmetry, only polar fluid modes\footnote{We remind that
polar modes are those whose parity for a given spherical harmonic
number $l$ is $(-1)^{l}$ while axial modes have $(-1)^{l+1}$ for
parity.} can exist and the spectrum of axial modes is degenerate at
zero frequency. Moreover, using the relativistic generalization of the
anelastic approximation [{\it cf.} Eq.(\ref{e:ane_cons}), Paper I
where it was introduced and also the discussion in the Appendix], the
only hydrodynamical modes left in the nonrotating cold star are
low-frequency modes whose restoring force is gravity, the so-called
composition {\it g-}modes. Indeed, the main consequence of the
anelastic approximation is to filter out pressure modes (see the
Appendix). In addition, the relativistic Cowling approximation
prevents us from having {\it w-}modes (some of which could have been
axial), which are mainly oscillations of the spacetime itself.\\

 Gravity modes of NSs have already been the subject of several works,
even if they were first thought to be degenerate at zero frequency due
to beta equilibrium (Thorne, \cite{thorne1969}). Thus, gravity modes
that exist because of the presence of temperature gradients in warm
NSs were studied by Mc Dermott {\it et al.} \cite{mcdermott83}, who
introduced the relativistic Cowling approximation. They found that the
thermal {\it g}-modes were concentrated into a thin layer close to the
surface, as was the case in the analysis of Finn \cite{finn1987} on
gravity modes linked to discrete changes in composition of the
crust. The first study considering Newtonian gravity modes associated
with smooth composition gradients in the core of NSs is that of
Reisenegger \& Goldreich \cite{reisenegger92}. However, they neglect
the effect of the strong interaction, which is inappropriate since
nucleons are far from Fermi gases [see for instance
Eq.(\ref{e:pal1})]. Later, Lai \cite{lai99} studied the CFS
instability of composition gravity modes using EOSs that include the
strong interaction (see \cite{lai94}), but in Newtonian rotating
NSs. While Yoshida \& Lee \cite{yoshida02} looked at the {\it r}-modes
of relativistic stars with gravity modes, yet they mainly focussed on
modes due to temperature gradients, and only dealt in a very
approximative way with modes due to composition gradients. Going
farther in temperature, gravity modes due to discontinuity of density
in warm nonrotating relativistic NSs were investigated by Miniutti
{\it et al.} \cite{miniutti03}, whereas even more recently gravity
modes due to temperature and composition gradients in hot rotating
relativistic NSs has been analysed by Ferrari {\it et al.}
\cite{ferrari04}, restricting themselves to polar modes. On the other
hand, the situation of gravity modes in very cold (superfluid) NSs was
first studied by Lee \cite{lee95}, who realised that there was no
gravity modes in models with two fluids. This was later confirmed by
Andersson \& Comer \cite{andersson01} and Prix \& Rieutord
\cite{prix02}, and is now considered as a crucial feature of superfluidity to prove
its existence in NSs through the observation of gravitational wave signals.\\

  Nevertheless, as already explained, we shall deal here only with
  modes of not too cold neither too hot NSs, and the main
  characteristics of the relevant composition gravity modes resulting
  from the previously mentioned studies are
\begin{description}
\item[(i)] a weak coupling to the crust and the independence with
respect to the transition radius between the core and the inner-crust
(\cite{reisenegger92},\cite{lai99}). Following this result, we decided
that it was appropriate to neglect the coupling between the core
gravity modes and the crust modes, just giving a boundary condition
for the modes at the transition density. Notice that in Paper I we
found that using either free surface boundary condition or rigid crust
boundary condition was not a key issue in the linear approach to
inertial modes. Hence, in all the following, only modes with the
boundary condition of a rigid crust (null radial velocity) at a given
density are calculated. This density is of course the chosen density
for the cutoff of the PAL EOS (see Section \ref{sec:bgm}), {\it i.e.}
it is either half the saturation density or almost two thirds of it.
The influence of the cutoff on the mode shall be discussed in the next
subsection;
\item[(ii)] the growth time of their gravity driven instability, which
is much longer than the time of viscous damping, makes the {\it
g-}modes irrelevant for the production of gravitational waves (at
least for cold NSs, see \cite{ferrari04}). This comes from their weak
coupling to the gravitational field. This consideration is pointless
for a nonrotating star, but it explains why in the next section we
shall only deal with the current quadrupole that makes $m\,=\,2$
inertial modes unstable;
\item[(iii)] their frequencies, which are roughly between $50$ and $400$ Hz
for a $M\,=\,1.4\,M_\odot$ star (\cite{reisenegger92},\cite{lai99},\cite{ferrari04}),
decrease for a slightly decreasing size of the core.
\end{description}

The latter point turns out to be easy to understand from the usual
dispersion law for {\it g-}modes that can be reached in the Newtonian
case using the WKB approach (see the Appendix)
\be \label{e:wkb}
w^2\,\sim\,\frac{l(l+1)}{(k\,r)^2\,+\,l(l+1)}\,{\cal N}^2\,,
\ee
where $w$ is the frequency, $l$ the quantum number of the
decomposition in spherical harmonics, $r$ the radial coordinate, $k$
the radial component of the wave vector and ${\cal N}^2$ the
Brunt-V\"ais\"al\"a frequency defined as
\be \label{e:brunt2}
{\cal N}^2\,\df\,\frac{1}{n\,P}\,\left|\nb P\right|^2\,
(\gamma_{\beta}^{-1}\,-\,\gamma_F^{-1})\,. 
\ee

 Notice that these $\gamma$ are the Newtonian equivalent of the relativistic
gamma coefficients\footnote{The difference between relativistic and
Newtonian gamma coefficients is that the baryonic density $n_b$ has to
be replaced with the mass density $n$ in the Newtonian case.}
introduced in Section \ref{sec:perthydro}, and also that in the Appendix this
Newtonian formula is written with the sound velocities which are easily related
to those coefficients.\\

 Looking at Equation (\ref{e:wkb}), we see that for a given
$l$, to make the radius of the star a little smaller (which means
to increase the value of $k$ without changing ${\cal N}$) implies a
decrease of $w$ in accordance with point {\bf (iii)} above.\\

\begin{figure}
\begin{center}
\includegraphics[height=8.6cm,angle=-90]
{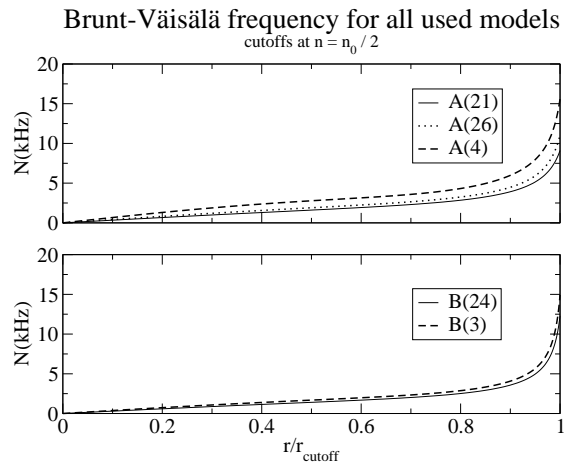}
\caption[]{\label{fig:brunt} Brunt-V\"ais\"al\"a frequency for all the
models used in this work up to the radius for which the density is
half the saturation density. Note that since the frequency vanishes at
the center of the star (null gradient of pressure) and monotonously
increases with the radius, gravity modes will have an allowed
propagation zone with a lower limit.}
\end{center}
\end{figure}

\begin{figure}
\begin{center}
\includegraphics[height=8.6cm,angle=-90]
{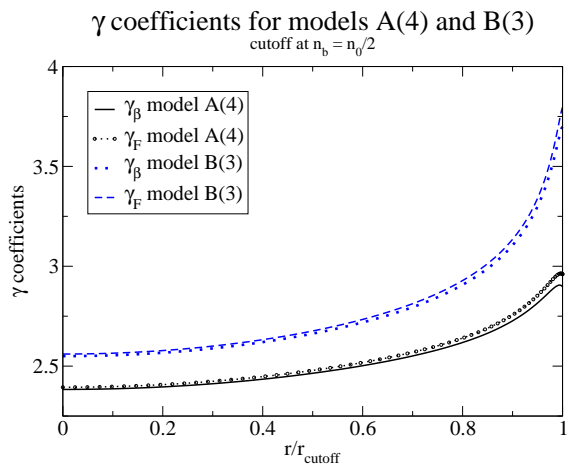}
\caption[]{\label{fig:gam} For the EOSs A and B, $\gamma$ coefficients
as functions of the radius (in units of the cutoff radii) in the stars
with maximal central densities in order to scan the wider ranges as
possible. Notice that the frozen coefficients are always higher than
the coefficients at equilibrium, which is a necessary condition for
stability.}
\end{center}
\end{figure}

   Since ${\cal N}$ depends on the background star, the variation of the
frequencies for changes of mass or of EOS are more subtle. To have an
insight on this in our case, it is useful to look at the relativistic
equivalent of Eq.(\ref{e:brunt2})
\be \label{e:bruntrel}
{\cal N}^2\,\df\,\frac{c^2}{P\,(\rho\,+\,P)}\left|\nb
P \right|^ 2\,(\gamma_{\beta}^{-1}\,-\,\gamma_F^{-1})\,,
\ee
where we restored explicitly the velocity of light $c$. The values of
this frequency $N$ for our models are depicted in Figure
\ref{fig:brunt}, whereas Fig.\ref{fig:gam} shows, as an indication,
the gamma coefficients for the models with maximal central densities
[A(4) and B(3)]. In Fig.\ref{fig:brunt}, we see that the
Brunt-V\"ais\"al\"a frequency (that is an upper-bound for the allowed
frequencies) seems to be higher for softer EOSs [comparison between
models A(4) and B(3) that have the same mass], and increases with the
mass for a given stiffness. In the previous studies, this point did
not seem very clear, since Reisenegger \& Goldreich found that the
frequency of gravity modes increases with mass (which is consistent
with the increase of the Brunt-V\"ais\"al\"a frequency), but the
result of Lai with fixed mass was that an apparently softer EOS
(smaller radius) leads to a smaller Brunt-V\"ais\"al\"a frequency. Yet,
it is worth pointing out that the properties of a NS depend not only on
the compression modulus, but also on the symmetry energy. Since
Lai used two completely different EOSs, his results are
probably not an indication of the influence of the stiffness on the
Brunt-V\"ais\"al\"a frequency nor on the gravity mode frequency.
This issue will be addressed in the next section.

\subsection{Test of the code and gravity modes} \label{sec:gmod2}

 The first test that we did was to take one of the background
 stars presented in Section \ref{sec:bgm} [model A(21)] and perturb it with a
 small $m=2$ Lagrangian perturbation of density
\begin{eqnarray}
\label{eq:denspert2}
\Dt n_b\,=&\,r^3\left(1\,-\,r^2\right)\,\sin(\th)^2\,\cos(\th)\,\nonumber\\
&\times\,\left[\cos(2\,\phi)\,+\,\sin(2\,\phi)\right]\,.
\end{eqnarray}

\begin{figure}
\begin{center}
\includegraphics[height=8.6cm,angle=-90]{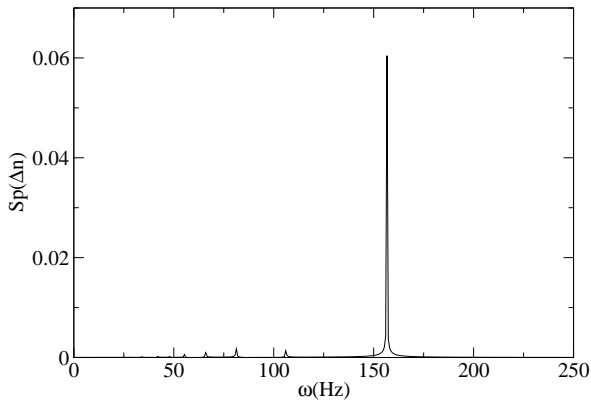}
\caption[]{\label{fig:specdens} Fourier power spectrum of the time
evolution of a Lagrangian perturbation of density on the equator
for $m\,=\,2$ initial data. The total physical duration is 1.5 s.
See also Figure \ref{fig:specdens2} for a logarithmic representation.}
\end{center}
\end{figure}

\begin{figure}
\begin{center}
\includegraphics[height=8.6cm,angle=-90]
{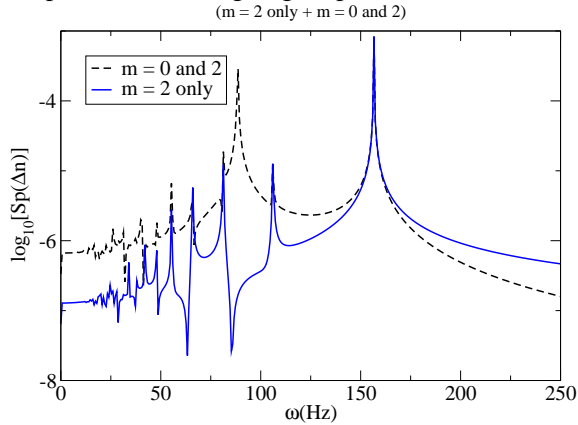}
\caption[]{\label{fig:specdens2} Same Fourier power spectrum as in
Figure \ref{fig:specdens} but in logarithmic scale. In addition is
pictured the spectrum for another calculation that corresponds to a
mixture of $m\,=\,0$ and $m\,=\,2$ as initial data. The agreement
concerning the values of the frequencies of the modes is very good and
a $l=1$ mode, which only exists for $m=0$, is the main difference between the two spectra.}
\end{center}
\end{figure}

 The result and the stability of the code are illustrated in Figures \ref{fig:specdens} and
 \ref{fig:specdens2} that depict the Fourier spectra of the evolution.
Figure \ref{fig:specdens2} is in logarithmic scale as will be
 every similar figure in the following: in this way more modes are visible.
 With this model, the higher frequency gravity mode has a frequency
 around $150$ Hz. This order of magnitude is in agreement with
 previous results (\cite{reisenegger92},\cite{lai99}), but we shall
 discuss their values and how they depend on the background star in
 the following.\\

   In Figure \ref{fig:specdens2}, we also put another calculation done
 with the same background star, but changing the initial data. Instead
 of using initial data whose angular decomposition corresponds to one
 single azimuthal number ($m\,=\,2$), we tried initial data with
 $m\,=\,0$ and $m\,=\,2$ mixed:
\begin{eqnarray}
\label{eq:denspert02}
\Dt n_b\,=&\,r^3\left(1\,-\,r^2\right)\,\sin(\th)^2\,\cos(\th)\,\nonumber\\
&\times\,\left[\cos(2\,\phi)\,+\,\sin(2\,\phi)\,-\,1\right]\,.
\end{eqnarray}

 It can be shown that while the first data were only composed of the
 $(l=3, m=\pm2)$ associated Legendre polynomials, the second data
 additionally contain $(l=1, m=0)$. Hence, this figure shows that, as
 expected in a spherical star (nonrotating and without anisotropic
 physics), the modes that can be seen do not depend on the azimuthal
 numbers present in the angular decomposition. Exception done of the
 fact that for a given value of $m$ only modes with $l>|m|$ can
 exist. This is the reason why for $(m=0)$ an additional mode with the
 frequency $\sim88.8$ (Hz) (plus others with very small frequencies) appears.\\

\begin{figure*}
\begin{center}
\includegraphics[height=15.cm,angle=-90]{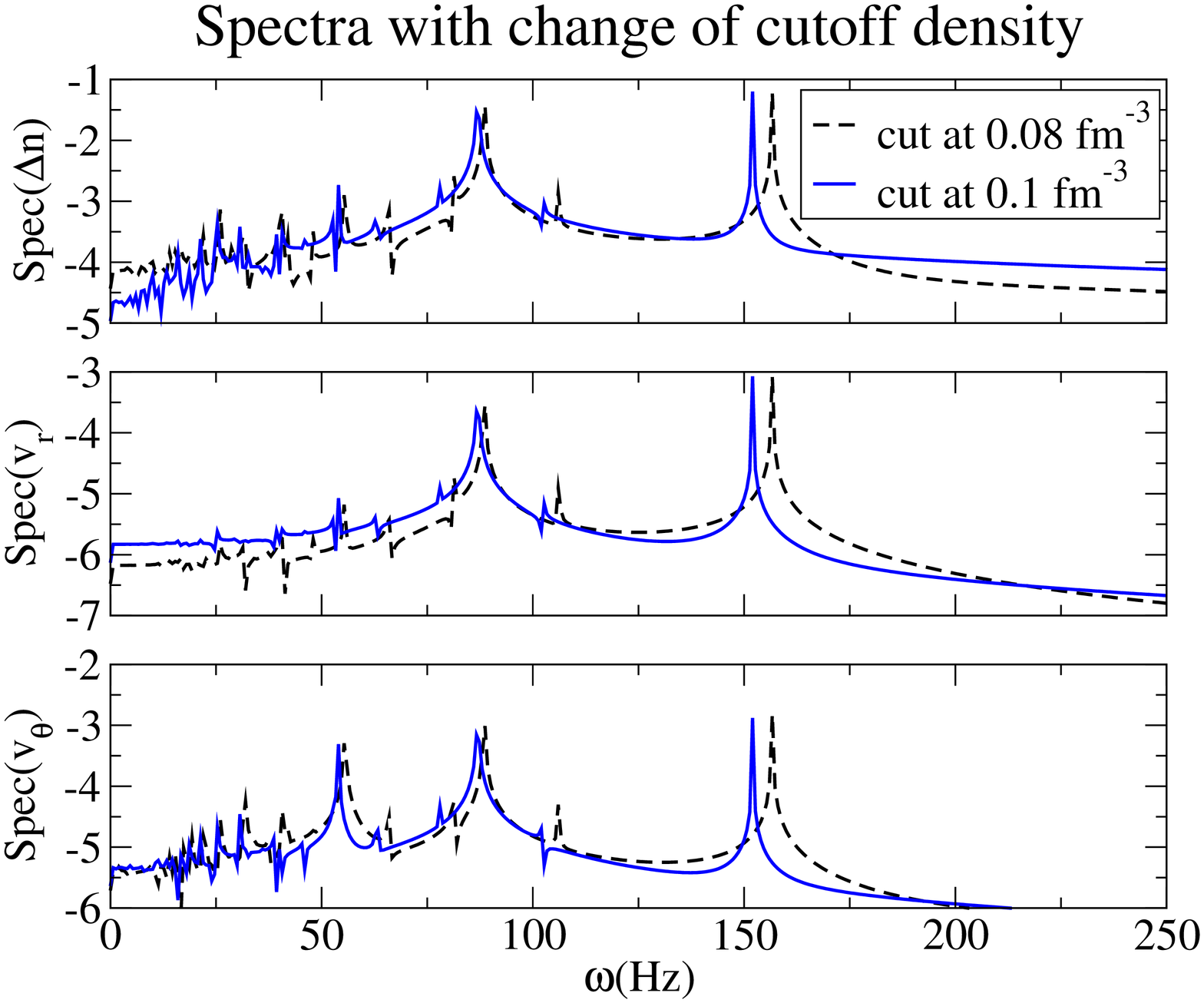}
\caption[]{\label{fig:speccut} Fourier power spectra of Lagrangian density
perturbation, radial and $\th$ components of velocity for a
star with two different values of the density at which is done the
cutoff. As expected for gravity modes, the frequencies decrease when the size
of the star is made smaller by increasing the value of the cutoff
density. Yet, the difference is quite small.}
\end{center}
\end{figure*}

 Since our main goal is to study inertial modes with $m\,=\,2$ (see
 Section \ref{sec:rotmod}) that are the most interesting from the
 gravitational waves point of view, we shall restrict ourselves only
 to $m\,=\,2$ initial data as soon as we shall deal with rotating
 stars. But for the time being, we shall keep the mixed initial data
 in the study of gravity modes without rotation. The next point that
 we shall discuss is the influence of the cutoff density. Following
 the already mentioned works on gravity modes in NSs, we expect that
 this cutoff should hardly have an influence on the spectrum, the main
 influence being linked to the change in the size of the core. This
 point is clarified by Figure \ref{fig:speccut} that shows, for the
 same background star as before, spectra corresponding to the time
 evolution of the Lagrangian perturbation of (baryonic) density, but
 also of the radial and $\th$ components of the (Eulerian perturbation
 of) velocity for two cutoff values: $0.08$ or $0.1$ fm$^{-3}$. Note
 that these spectra are reached with evolutions done at different
 places inside the star. With Figure \ref{fig:speccut}, we are able to
 verify that to cut at a higher density (which means to make the star
 smaller) indeed decreases the frequency, but the change is quite
 small: some $3 \%$ in the worst case. Hence, we shall now restrict
 ourselves only to calculations done with a cutoff value of
 $0.1\,\rm{fm}^{-3}$. Moreover, we also see in this figure that there
 is a very good agreement between spectra reached from time evolutions
 of density or of a component of the velocity. In addition, the fact
 that not all gravity modes have a radial velocity (gravity modes can
 be either polar or axial) is illustrated, mainly with low-frequency
 modes.\\

\begin{figure*}
\begin{center}
\includegraphics[height=15.cm,angle=-90]{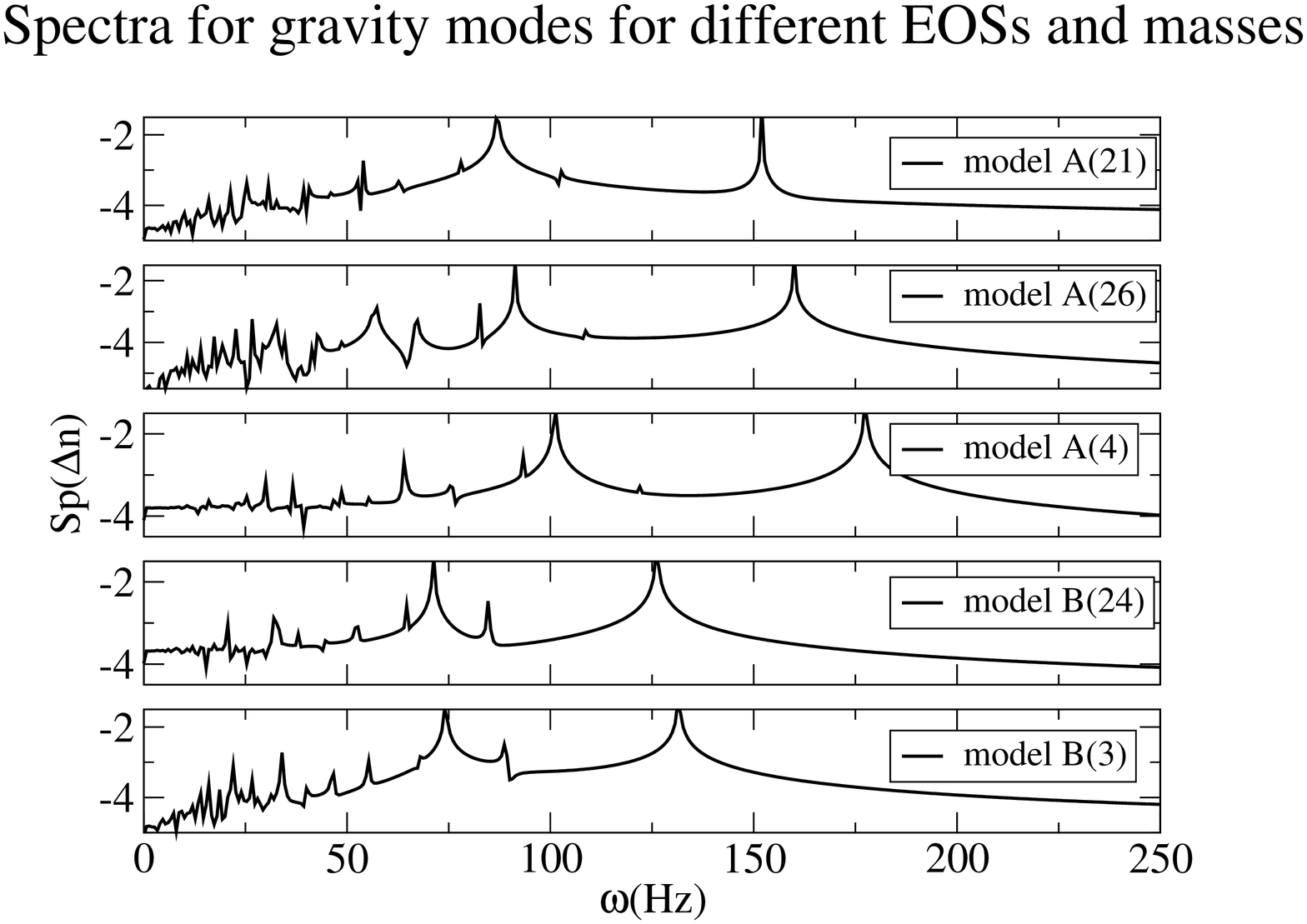}
\caption[]{\label{fig:specmass} Fourier power spectra of Lagrangian
density perturbation for stars with different masses and EOSs. All the
models used in this study are presented here.}
\end{center}
\end{figure*}

  The next step in our calculation was to look for the influence on
  gravity modes of the physical parameters that describe the
  background star. This is depicted in Figure \ref{fig:specmass}
  where we draw the frequency of the gravity modes for all models
  presented in Table \ref{tab:bg}. Thus, the three first graphs depict
  spectra for a relatively stiff matter (compression modulus of 180
  MeV) with increasing central density (from top to bottom) whereas the
  last two curves are for an EOS which is stiffer: compression modulus
  of 240 MeV. The first comment to make is that to increase the mass
  (the central density) for a given EOS shifts the spectra to higher
  frequencies. This result is in agreement with the increase of
  Brunt-V\"ais\"al\"a frequency found in Figure \ref{fig:brunt}.\\

 Next, comparison between the third and the fifth curves also gives a
  dependence in agreement with Figure \ref{fig:brunt}. These two
  curves indeed correspond to stars with the same masses, but with
  different stiffness. Hence, we verify that stiffer EOSs [model B(3)]
  lead to lower frequencies, even if they imply smaller stars. The
  radius rather than the mass seems to be the dominant parameter to
  determine the value of the highest gravity mode. This is made more
  evident when we look at the second and the third curves [models
  A(26) and A(4)] and at the fourth and the fifth curves [models B(24)
  and B(3)]. Indeed, the difference between the mass of A(26) and A(4)
  is about 0.25 solar masses for a difference in the frequency of the
  highest gravity mode of 17 Hz (160 to 177 Hz), while for B(24) and
  B(3), the difference between the masses is around 0.2 solar masses
  with only 5 Hz between the frequencies (126 to 131 Hz). However, as
  we can see in Table \ref{tab:bg}, the difference between the radii
  is of 0.4 km for model A and only 0.03 km for model B. In fact, this
  result only illustrates the same as in Figure \ref{fig:brunt}:
  matter added to a star with a stiff EOS hardly changes the
  radius. This is the reason why the Brunt-V\"ais\"al\"a frequency was changed between
  models A(26) and A(4), but was hardly modified between B(24) and
  B(3). Yet, as we shall see now, when the star is rotating it is more
  tricky to identify modes just by looking at a spectrum.

\section{Inertial modes and composition gravity modes in rotating
neutron stars} \label{sec:rotmod}

\subsection{Modes in rotating stars}

  As explained in the Introduction, modes of rotating relativistic
stars have the peculiarity that some of them can be driven unstable by
their coupling to gravitational waves. In fact, this CFS
(\cite{chandra70},\cite{friedman78}) instability is generic, and only
insufficiently high angular velocity of a star or physical phenomena
like viscosity can prevent some oscillations of being
unstable. Furthermore, Friedman \& Schutz \cite{friedman78} proved
that the instability appears when an initially retrograde mode becomes
prograde (as seen by an inertial observer). To explain how we should
be able to ``detect'' this occurrence in the present work (in which we
only look at spectra with positive frequencies), it is worth to start
with a brief summary of the main differences between modes of rotating
and nonrotating stars.\\

  The first obvious effect of introducing rotation in the Euler
equation for the oscillations of a star is to allow the appearance of
modes restored by the Coriolis force (the so-called inertial
modes). But rotation also breaks the degeneracy at zero frequency of
purely axial modes. This leads to the possible existence of purely
axial inertial modes, the so-called {\it r}-modes, while it also
possibly couples together axial and polar modes. Another effect of
rotation, on the modes that already exist in the nonrotating star, is
the splitting of their frequencies. Thus, frequencies measured in the
inertial $w_i$ and rotating $w_r$ frames are linked by the relation
\be
w_i\,=\,w_r\,-\,m\,\omg\,,
\ee
where $m$ is the azimuthal number and $\omg$ the angular velocity of
the star (or of the rotating frame). Since we now restrict ourselves
to modes with $|m|\,=\,2$ (the most interesting for GW), this
splitting should appear in our spectra \underline{calculated in the
inertial frame} as the replacement of any mode (with frequency $w_0$
in the nonrotating case) by a pair of modes with frequencies
$w_r\,+\,2\,\omg$ and $w_r\,-\,2\,\omg$, with $w_r\,\sim\,w_0$.
Notice that the small difference between $w_r$ and $w_0$ is linked
to the influence of rotation on the structure of the star and on the
modes frequencies, a difference which should remain quite small since
we keep here only terms linear in $\omg$ (slow rotation
approximation).\\

 For such a pair of frequencies, the mode whose positive frequency decreases
with increasing $\omg$ is obviously the retrograde mode. Hence, in
Section \ref{sec:rotnb} where we shall have gravity modes in rotating
stars, a possibly unstable gravity mode will be detected when in the
spectra (for increasing $\omg$) its frequency have reached zero
and then started to grow again. But as mentioned in Sections
\ref{sec:visco} and \ref{sec:gmod}, gravity modes of cold NSs are not good
candidates for the emission of gravitational waves due to the CFS
instability. They are weakly coupled to the gravitational field and we
shall then not pay too much attention to their potential instability.\\

 The situation is quite different for inertial
modes. Indeed, restored (in the Newtonian limit) by the Coriolis
force, they only exist for rotating stars and have frequencies
proportional to the star's angular velocity $\omg$. Thus, the
Newtonian frequency of a purely axial inertial mode is, in the
rotating frame,
\be 
w_r\,=\,\frac{2\,m\,\omg}{l(l+1)}\,,
\ee
which gives the always negative frequency (which means an always
prograde mode) in the inertial frame
\be \label{e:rmn}
w_i\,=\,-\,m\,\omg\,\frac{(l+2)(l-1)}{l(l+1)}\,=\,-\,m\,
\l(\omg\,-\,\frac{2\,\omg}{l(l+1)}\r)\,.
\ee

  This is the reason why, as mentioned in the Introduction, some
inertial modes are unstable whatever the angular velocity of the
background star is. Moreover, it can be shown (\cite{friedman01},
\cite{andersson03a}, \cite{kokkotas03}) that depending on the assumed
motion (whether or not the perturbations are adiabatic), the properties
of inertial modes change. Thus, a Newtonian barotropic star has a
spectrum in which only axial modes ({\it r}-modes) subsist for
spherical harmonics that satisfy $l\,=\,m$, whereas
nonbarotropic stars admit axial modes for any combination of $l$ and
$m$.\\

 In the relativistic case, the situation is a bit different, starting
with the fact that no pure axial modes persist for barotropic stars as
proven by Lockitch {\it et al.} \cite{lockitch01} (see also Paper
I). For nonbarotropic relativistic stars, the problem was quite
controversial during some time and it is still not so clear. Indeed, in the
relativistic framework, purely axial modes of nonbarotropic slowly
rotating stars are expected to satisfy the Kojima's master equation
\cite{kojima98}. This equation contains some perturbations of the
metric, but it was shown \cite{kokkotas03} that the main features of
{\it r}-modes do not change with the relativistic Cowling
approximation. Moreover, with this approximation, due to the existence
of the frame-dragging effect, this equation easily leads to the
relativistic equivalent of Eq.(\ref{e:rmn})
\be \label{eq:disprel}
w_i\,=\,-\,m\,\left(\omg\,-\,\frac{2\,\varpi}{l(l+1)}\right)\,,
\ee
in which
\be
\varpi\,\df\,\omg\,-\,N_{\ph}\,,
\ee
where $N_{\ph}$ is the third component of the shift vector defined in
the metric (\ref{e:metric}). Since $\varpi$ is a function of $r$ (and
only of $r$ in the slow rotation approximation), Eq.(\ref{eq:disprel})
implies a ``continuous spectrum'' (see \cite{beyer99} for a proper
demonstration). This made unclear for some time the existence of
discrete {\it r}-modes for certain relativistic stars (for more detail
see the review by Kokkotas \& Ruoff \cite{kokkotas03}). For instance,
Ruoff and Kokkotas (see references within \cite{kokkotas03}) found
that only for a very restricted range of polytropic stars was possible
to encounter purely axial inertial modes. Thus, the larger is the
polytropic index of a star, the lower its maximal compactness can be
for it to admit discrete {\it r-}modes in its spectrum.\\

 But it was also demonstrated \cite{lockitch01} that Kojima's master
equation becomes a singular eigenvalue problem if the frequency in
the rotating frame, $w_r\,\df\,w_i\,+\,m\,\omg$, verifies
\be \label{eq:range}
\frac{2\,m\,\varpi(0)}{l(l+1)}\,\le\,w_r\,\le\,
\frac{2\,m\,\varpi(R)}{l(l+1)}\,,
\ee
where $0$ is the null radius (center of the star) and $R$ the radius
at the surface. For this reason, it was claimed by Lockitch \&
Andersson \cite{lockitch02} that some ``boundary layer like'' approach
could be the way to properly solve this problem. Later Ruoff {\it et
al.} realized \cite{ruoff03} that to include a coupling between
$l$ terms of axial modes with $l\,\pm\,1$ terms of polar modes (and
respectively) could imply the existence of modes with discrete
frequencies, some of them possibly hidden in the continuous part of
the spectrum. However, they also found that this continuous spectrum was
hugely influenced by the number of $l$ they coupled together.\\

  Thus, the oscillation spectrum of a nonbarotropic rotating
relativistic neutron star is still an open problem that we shall deal
with in Section \ref{sec:rotnb}, in which modes of the background NSs
described in Table \ref{tab:bg} will be studied taking into account
the frozen composition. Hence, in order to check more easily if the
frequencies that will be displayed are inside or outside of the range
(\ref{eq:range}), we summarize, in Table \ref{tab:bornes}, their
limits for ($m\,=\,2\,,\,l\,=\,2\,\,\rm{or}\,\,3$) and for all our
models ({\it cf.} Table \ref{tab:bg}). But since these limits are defined in
the rotating frame while we work in the inertial frame, we put in this
table not directly $\varpi$ but $\hat{\varpi}_{lm}$ defined as
\be
\hat{\varpi}_{lm}\,\df\,\frac{1}{\omg}\,
\l(\frac{2\,m\,\varpi}{l(l+1)}\,-\,m\r)\,.
\ee

 Finally, to better recognize the influence of the assumption of
frozen composition, we shall first focus on inertial modes of those
stars with the hypothesis of barotropic EOSs.


\begin{table*}
\centering
\caption[]{Dimensionless limits of the ranges of continuous spectra of
some purely axial inertial modes as seen in the inertial frame for all
models presented in Table \ref{tab:bg}. In order to use these limits
for arbitrary angular velocity $\omg$, the dimensionless $\hat{\varpi}_{lm}$
is introduced.}

\label{tab:bornes}
\begin{tabular}{ c |  c  | c | c || c | c  }
  \hline
  \hline
 EOS  & Model A(21)  & Model A(26) &  Model A(4) & Model B(24) & Model B(3)  \\
  \hline
$n_\mathrm{surf}$ (fm$^{-3}$) & 0.08 \, 0.1& 0.08 \, 0.1
& 0.08 \, 0.1& 0.08 \, 0.1& 0.08 \, 0.1\\
\hline
$\hat{\varpi}(0)_{22}$ &  -1.602 \, -1.600 & -1.643 \, -1.642 
& -1.731 \, -1.733  & -1.631 \, -1.631 & -1.681 \, -1.675\\
$\hat{\varpi}(0)_{32}$ &  -1.801 \, -1.800 & -1.821 \, -1.821
& -1.865 \, -1.866  & -1.816 \, -1.816 & -1.841 \, -1.837\\
\hline
$\hat{\varpi}(R)_{22}$ &  -1.422 \, -1.423 & -1.436 \, -1.438
& -1.468 \, -1.473  & -1.436 \, -1.439 & -1.459 \, -1.455\\
$\hat{\varpi}(R)_{32}$ &  -1.711 \, -1.711 & -1.718 \, -1.719
& -1.734 \, -1.736  & -1.718 \, -1.719 & -1.729 \, -1.727\\
\hline
  \end{tabular}
\end{table*}


\subsection{Modes of relativistic rotating barotropic stars}

\begin{figure*}
\begin{center}
\includegraphics[height=15.cm,angle=-90]{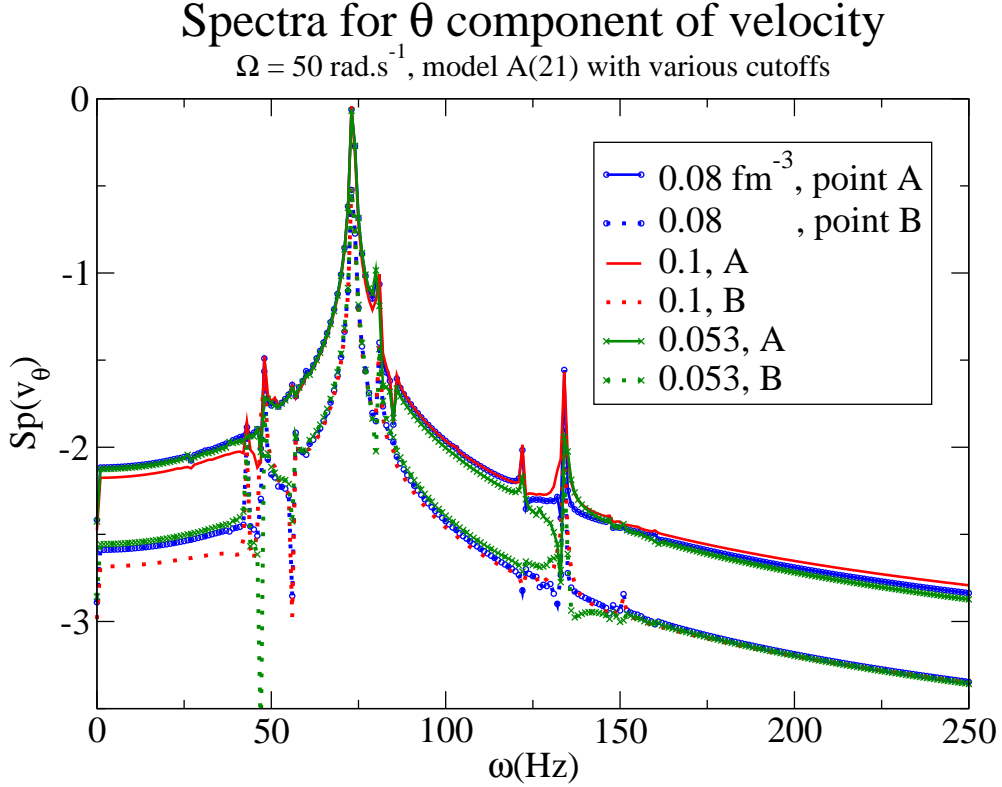}
\caption[]{\label{fig:specrmcutvt} Fourier power spectra of the $\th$
component of velocity for inertial modes in a barotropic star with
various values of the cutoff density. The spectra are calculated at 2
different positions inside the star illustrating the fact that they
are discrete modes. Numerous inertial modes can be identified.}
\end{center}
\end{figure*}

 The first thing that we shall look at is the influence of our cutoff
density on inertial modes. To explore this, we did several time
evolutions with the Newtonian $l\,=\,m\,=\,2$ {\it
r-}mode as initial data (see Paper I). This purely axial mode was evolved in
all backgrounds with different angular velocities of the star, and a
typical result is illustrated by Fig.\ref{fig:specrmcutvt}. This
figure depicts the spectra of the $\th$ component of the velocity for
a star of model A(21) with an angular velocity of $50$ rad.s$^{-1}$. Three
different values for the cutoff density are used: 0.053, 0.08 or
\mbox{0.1 fm$^{-3}$}, that is to say more or less $1/3$, $1/2$ or
$2/3$ of the saturation density $n_0$. Moreover, in order to test the
possible appearance of a continuous spectrum, we draw the spectra
calculated at 2 different positions within the star. Thus, the first
obvious result is that for this model at that precise angular
velocity, we see a discrete spectrum, with the
\mbox{$l\,=\,m\,=\,2$} inertial mode (the higher peak in the spectrum)
that seems to have a frequency around $73$ Hz. This value corresponds to a
ratio between the relativistic and Newtonian frequencies in the
rotating frame which is around $0.82$, in quite good agreement with
the results of Lockitch {\it et al.} \cite{lockitch03} for that
compactness. In addition, various inertial modes can be seen that can
be identified as the relativistic counterparts of the inertial modes
found by Lockitch \& Friedman \cite{lockitch99}, but since it was
shown \cite{lockitch03} that none of these modes are relevant for GW
with the CFS instability, we shall not describe them in detail.\\

\begin{figure}
\begin{center}
\includegraphics[height=8.6cm,angle=-90]{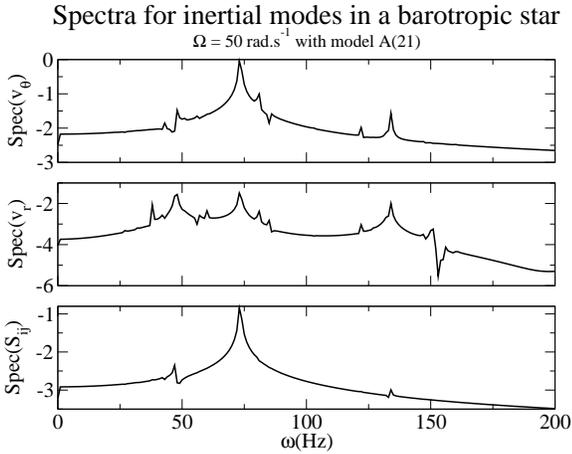}
\caption[]{\label{fig:rma21vtvrs} Fourier power spectra of the $\th$
and $r$ components of velocity and also of one of the 2 independent
components of the current quadrupole tensor $S_{ij}$ for inertial
modes in a barotropic star with cutoff density \mbox{$\sim 0.1$
fm$^{-3}$}. In addition to the $l\,=\,m\,=\,2$ inertial modes, others
can be seen.}
\end{center}
\end{figure}

 Moreover, as can be seen, the influence of the cutoff density on the
 spectrum of inertial modes is even smaller than for gravity
 modes. This can probably be understood from the fact that starting
 with purely axial initial data we reach a radial velocity smaller
 than in the case of the {\it g-}modes. This last statement is
 supported by the comparison between Figures \ref{fig:speccut} and
 \ref{fig:rma21vtvrs}. Both of these figures depicts spectra for time
 evolutions of $m\,=\,2$ modes in a star of model A(21). However, for
 the first one, the spectra of the $\th$ and $r$ components of the
 velocity correspond to gravity modes in a nonrotating star [initial
 data identical to the Eq.(\ref{eq:denspert2}) and still without velocity],
 while for the second they correspond to the Newtonian inertial mode in a barotropic
 A(21) star with \mbox{$\omg\,=\,50$ rad.s$^{-1}$}. Notice that for
 both calculations the cutoff density is $0.1$ fm$^{-3}$. What can be
 seen is that the orders of magnitude of the $\th$ component and of
 the radial component for the main peak are almost the same for
 gravity modes (Fig.\ref{fig:speccut}), whereas in the case of
 inertial modes (Fig.\ref{fig:rma21vtvrs}) the radial velocity is
 more than one order of magnitude smaller than the $\th$ component.
This feature can of course also be explained by the fact that the
spectra of relativistic inertial modes
 mainly depend on the compactness of the star as we shall further
 verify. Notice in addition that, as expected, this inertial mode is
 not a purely axial mode.\\

  The last curve of Fig.\ref{fig:rma21vtvrs} is the spectrum of one
of the two independent components of the current quadrupole tensor
(see Paper I for the exact definition). Since this tensor dominates the
post-Newtonian reaction force \cite{blanchet02}, we shall use it as an
indicator of the fastest growing modes, for those that verify the CFS
criterion. This indicator tells us
that the already mentioned inertial mode actually seems to be the
most unstable one. However, even if fewer modes can be seen here than in the
spectra of the velocity's components, several other modes appear. Among
them, a peak around $47$ Hz but also a quite hidden one around $81$ Hz
that is probably the axial-led $l\,=\,4$ mode whose stability was
discussed by \cite{lockitch03}: it has indeed a frequency in the
rotating frame that is some $70\,\%$ of the frequency of the
$l\,=\,m\,=\,2$ mode, while in the Newtonian case, Lockitch {\it et al.}
\cite{lockitch03} found a ratio of $69\,\%$. Notice that this argument
relies on the assumption of ``almost similarity'' of the spectra
calculated in the rotating frame for different stars. Formulated in
another way, it just means that the spectra in the rotating frames mainly
depend on one global physical parameter, the compactness. This
assumption will be tested in the following. Anyway, this mode was
shown not to be relevant for GW, which is supported by its very low
presence (compared with the main peak) in the $S_{ij}$ spectrum.
It was more noticeable in the velocity's spectra.\\

\begin{figure*}
\begin{center}
\includegraphics[height=15.cm,angle=-90]{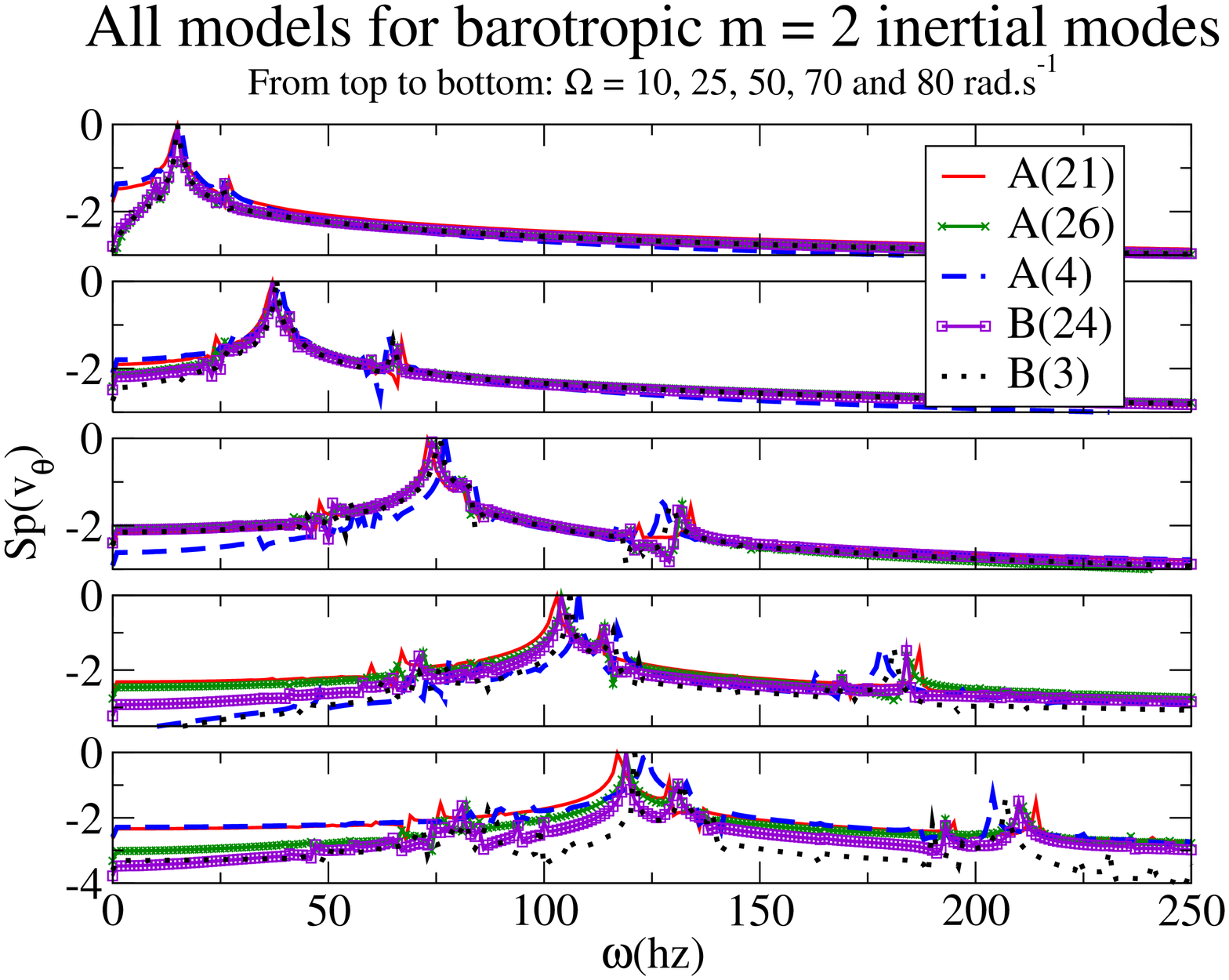}
\caption[]{\label{fig:rmallvth} Fourier power spectra of the $\th$
component of velocity for $m\,=\,2$ inertial modes using all our
background models with different angular velocity. The star is assumed
to be barotropic with a cutoff density at \mbox{$0.1$ fm$^{-3}$}.}
\end{center}
\end{figure*}

\begin{figure*}
\begin{center}
\includegraphics[height=15.cm,angle=-90]{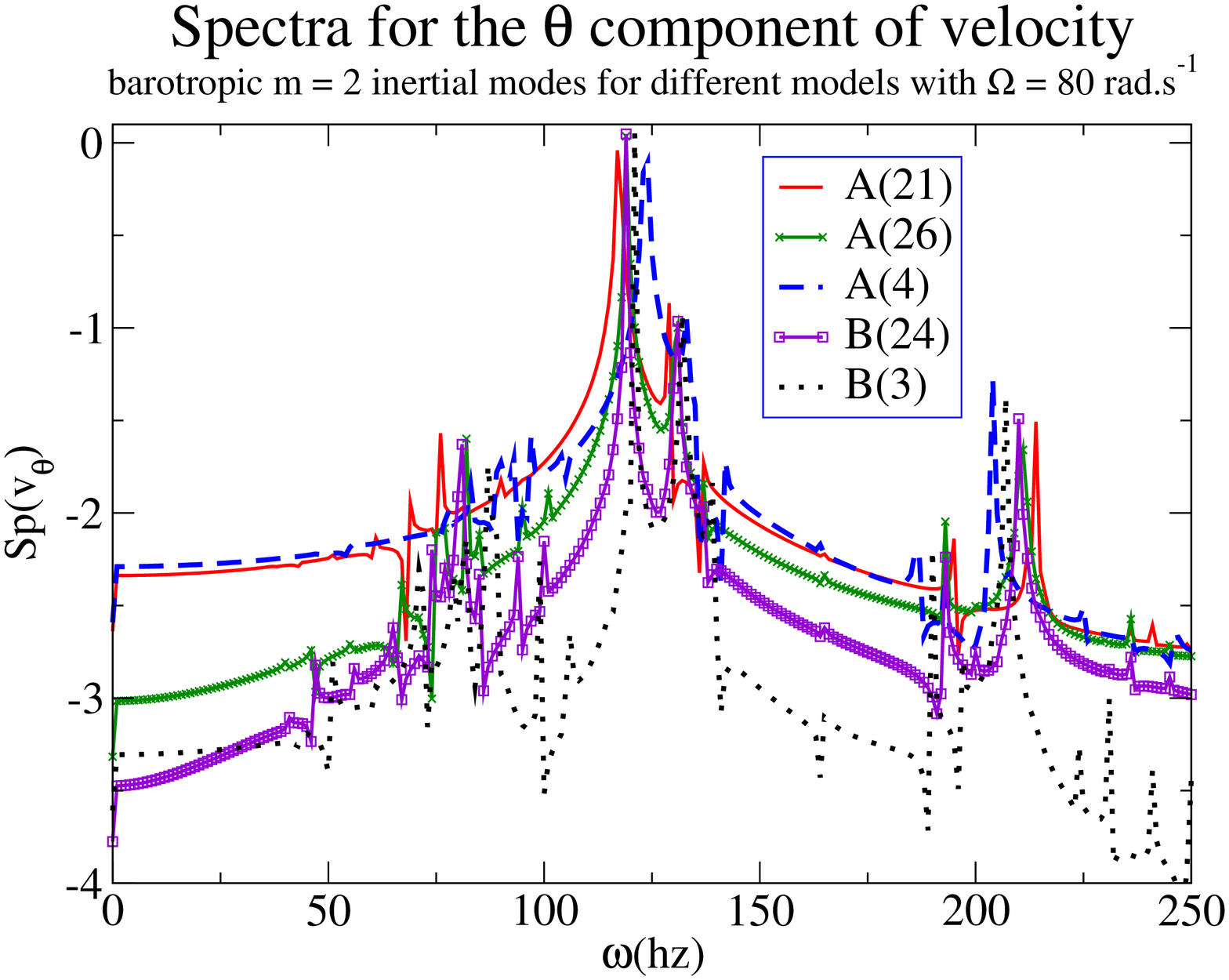}
\caption[]{\label{fig:rm80vth} Fourier power spectra of the $\th$
component of velocity for $m\,=\,2$ inertial modes using all our
background models for $\omg\,=\,80$ rad.s$^{-1}$. The star is assumed
to be barotropic with a cutoff density at \mbox{$0.1$ fm$^{-3}$}.}
\end{center}
\end{figure*}

  Now that we have analysed the effect of the cutoff density, we shall
be able to investigate the inertial modes for all our models. In
Fig.\ref{fig:rmallvth}, one can see the same kind of spectra for the
$\th$ component of velocity using all our background models and
changing the angular velocity from $10$ to $80$ rad.s$^{-1}$, while in
Fig.\ref{fig:rm80vth} only the curves for $\omg\,=\,80$ rad.s$^{-1}$
are depicted in order to better see the differences between different
stars. Thus, the first comment is that the influence of the background
star on the spectra is quite small. The higher resolution makes that
this is better seen in Fig.\ref{fig:rm80vth}, but already in
Fig.\ref{fig:rmallvth} one can observe a very small dispersion of the
frequencies. Moreover looking carefully, one can discern a cross
inside the highest square, {\it i.e.} the square denoting the
$l\,=\,2$ inertial mode for model B(24). Thus, models A(26) and B(24)
seem to have the same frequency for this mode, which supports the fact
that more than the EOS, the compactness of the star plays a key-role
in the difference between Newtonian and relativistic
frequencies. Indeed, as can be seen in Tab. \ref{tab:bg}, those stars
have almost exactly the same compactness: 0.189 and 0.191
respectively.\\

\begin{figure}
\begin{center}
\includegraphics[height=8.6cm,angle=-90]{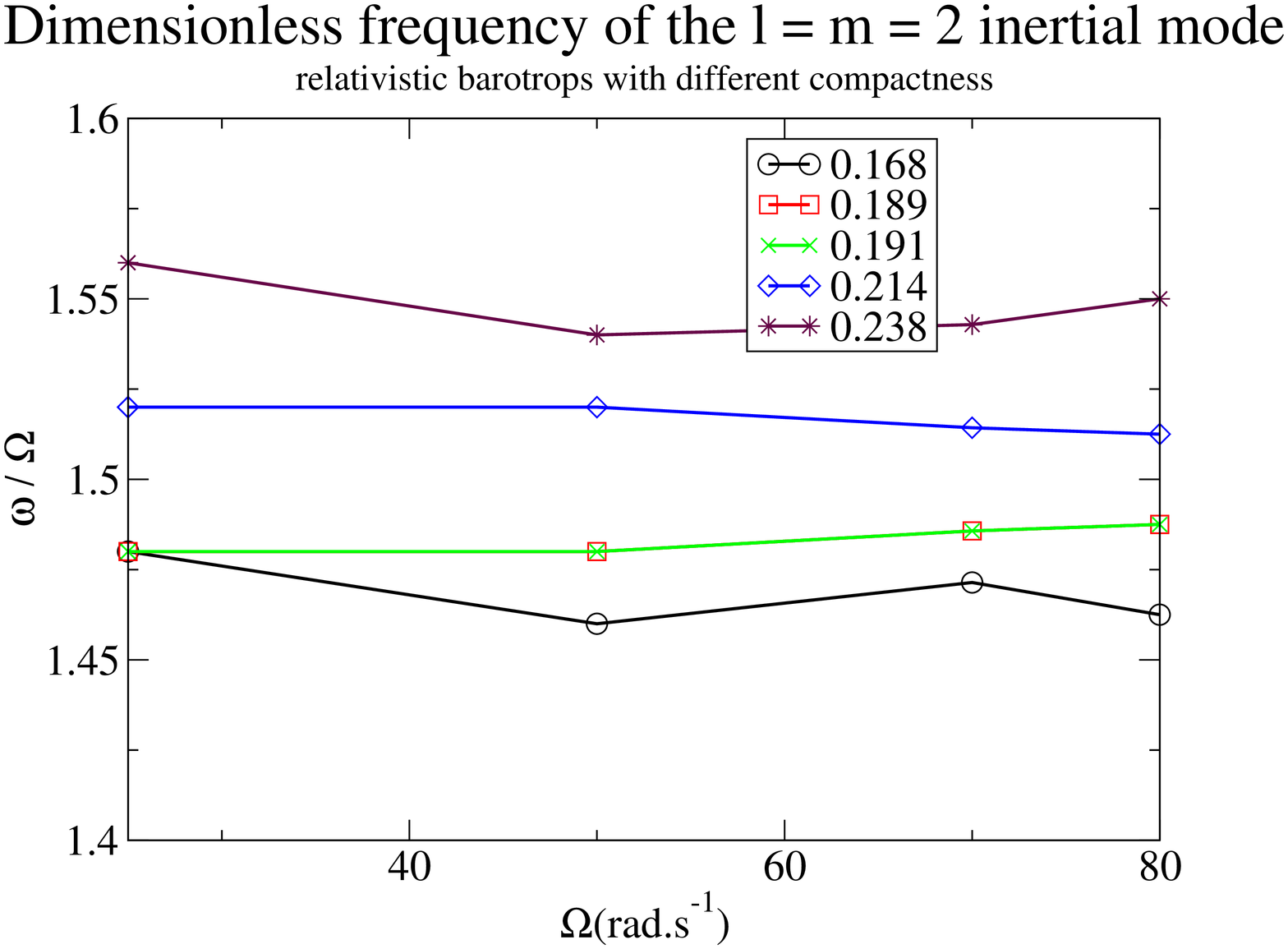}
\caption[]{\label{fig:lm2.baro} Dimensionless frequency versus angular
velocity of the star for \mbox{$l\,=\,m\,=\,2$} inertial mode in barotropic
stars of various compactness. Notice that due to the fact that evolutions
lasted $1$ s, the error bar is equal to $1\,\rm{rad.s}^{-1}/\omg$.}
\end{center}
\end{figure}

  Next, to test the dependence with respect to $\omg$ of the
  frequencies, we shall divide all of them by the angular velocity of
  the star. Since it is pointless for gravity modes, we shall restrict
  this analysis to the $l\,=\,2$ inertial mode, whose dimensionless
  frequency is illustrated versus the angular velocity for all
  compactness in Fig.\ref{fig:lm2.baro}. Nevertheless, before doing
  more comments about what turns out from this figure, it is worth to
  stress its limitation. Indeed, the version of the code used in this
  study works with a given time step of \mbox{$0.01$ ms}, and the total duration
  of the evolutions is always of the order of 1s, whilst the angular
  velocity changes. Hence, the error bars on frequencies expressed in
  Hz are of about 1 Hz, but for frequencies normalised with the angular
  velocity $\omg$, they are equal to $1\,\rm{rad.s}^{-1}/\omg$. Thus, for
  \mbox{$\omg\,=\,25$ rad.s$^{-1}$}, this is $4\,\%$ and this is
  less than $2\,\%$ for \mbox{$\omg\,>\,50$ rad.s$^{-1}$}. All of this
  explains why we did not put the data for \mbox{$\omg\,=\,10$
  rad.s$^{-1}$} that were completely meaningless.\\

 On this other hand, higher angular velocities of the star imply
  higher frequencies that give better resolution, which support the
  compactness dependence of the frequency as shown by
  Fig.\ref{fig:lm2.baro}. However, for all angular velocities, the
  frequencies of the $l\,=\,m\,=\,2$ mode for models with compactness
  $0.189$ and $0.191$ are the same\footnote{We verified that for
  higher frequencies inertial modes, the higher resolution enables us
  to make a distinction between them. We shall not discuss this since
  those modes are not relevant for gravitational waves
  emission.}. Yet, our main purpose was not to study inertial modes of
  barotropic stars, but to try to have an idea of the influence of the
  microphysical conditions, which implies nonbarotropicity, on the
  spectra. This is the reason why instead of working with time step
  and duration scaled by the inverse of $\omg$ (as we did in Paper I),
  we use here only ``physical scales'' that we shall also keep in the
  next section for the study of the modes of stratified rotating NSs.

\subsection{Modes of relativistic rotating nonbarotropic stars}
\label{sec:rotnb}

\subsubsection{Time evolution of a density perturbation}

   The first mentioned effect of rotation was the splitting of
noninertial modes. Our investigation of the pulsations of rotating
stratified relativistic NSs will begin with the illustration of
this phenomenon within the study of density perturbations in rotating
stratified NSs. To simplify, we shall first restrict ourselves to the
model A(21) with various angular velocities, but keeping the cutoff
density at $0.1$ fm$^{-3}$. In fact, since we have verified in the
previous sections its low influence, we shall use in the following
only that value of the cutoff density.\\

\begin{figure*}
\begin{center}
\includegraphics[height=15.cm,angle=-90]{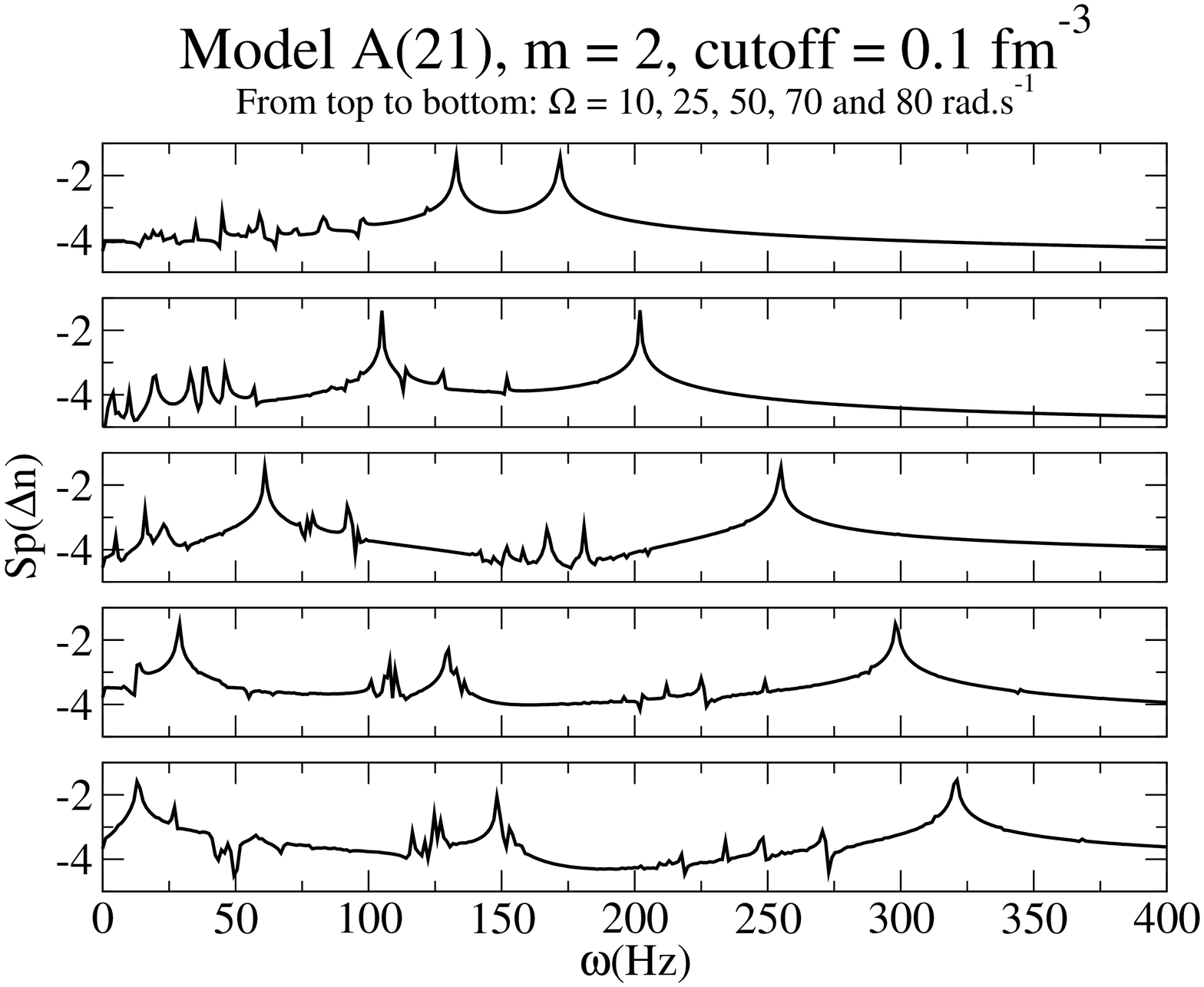}
\caption[]{\label{fig:dn.21.dn.om} Spectra of the Lagrangian
perturbation of density for several angular velocities. The star is
nonbarotropic and of model A(21), while the initial data are a
\mbox{$m\,=\,2$} density perturbation. See also
Figs. \ref{fig:vth.21.dn.om}, \ref{fig:vr.21.dn.om} and
\ref{fig:sij.21.dn.om}.}
\end{center}
\end{figure*}

\begin{figure*}
\begin{center}
\includegraphics[height=15.cm,angle=-90]{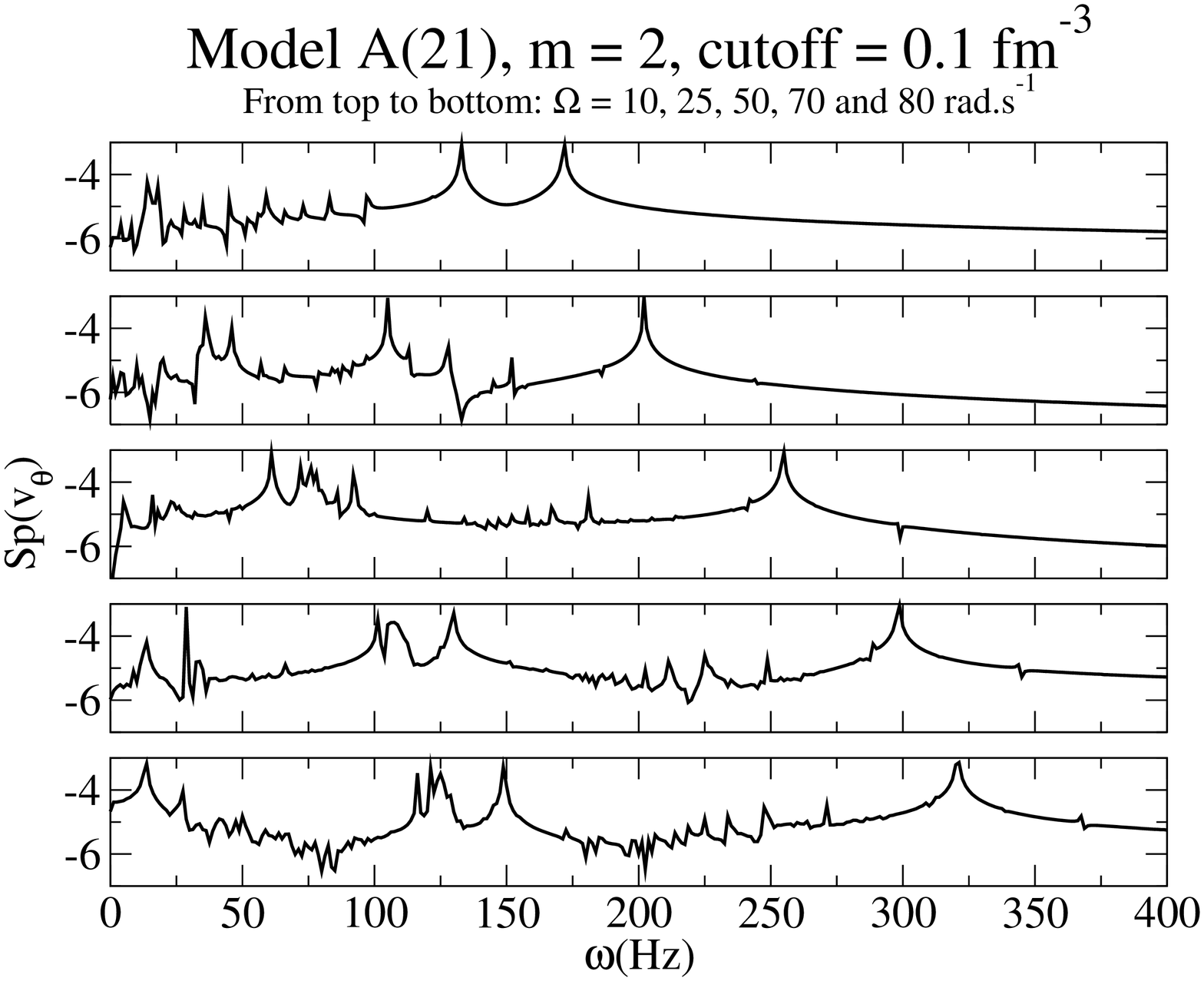}
\caption[]{\label{fig:vth.21.dn.om} Spectra of the $\th$ component of
velocity for several angular velocities. The star is nonbarotropic
and of model A(21), while the initial data are a \mbox{$m\,=\,2$}
density perturbation. See also Figs. \ref{fig:dn.21.dn.om},
\ref{fig:vr.21.dn.om} and \ref{fig:sij.21.dn.om}.}
\end{center}
\end{figure*}

\begin{figure*}
\begin{center}
\includegraphics[height=15.cm,angle=-90]{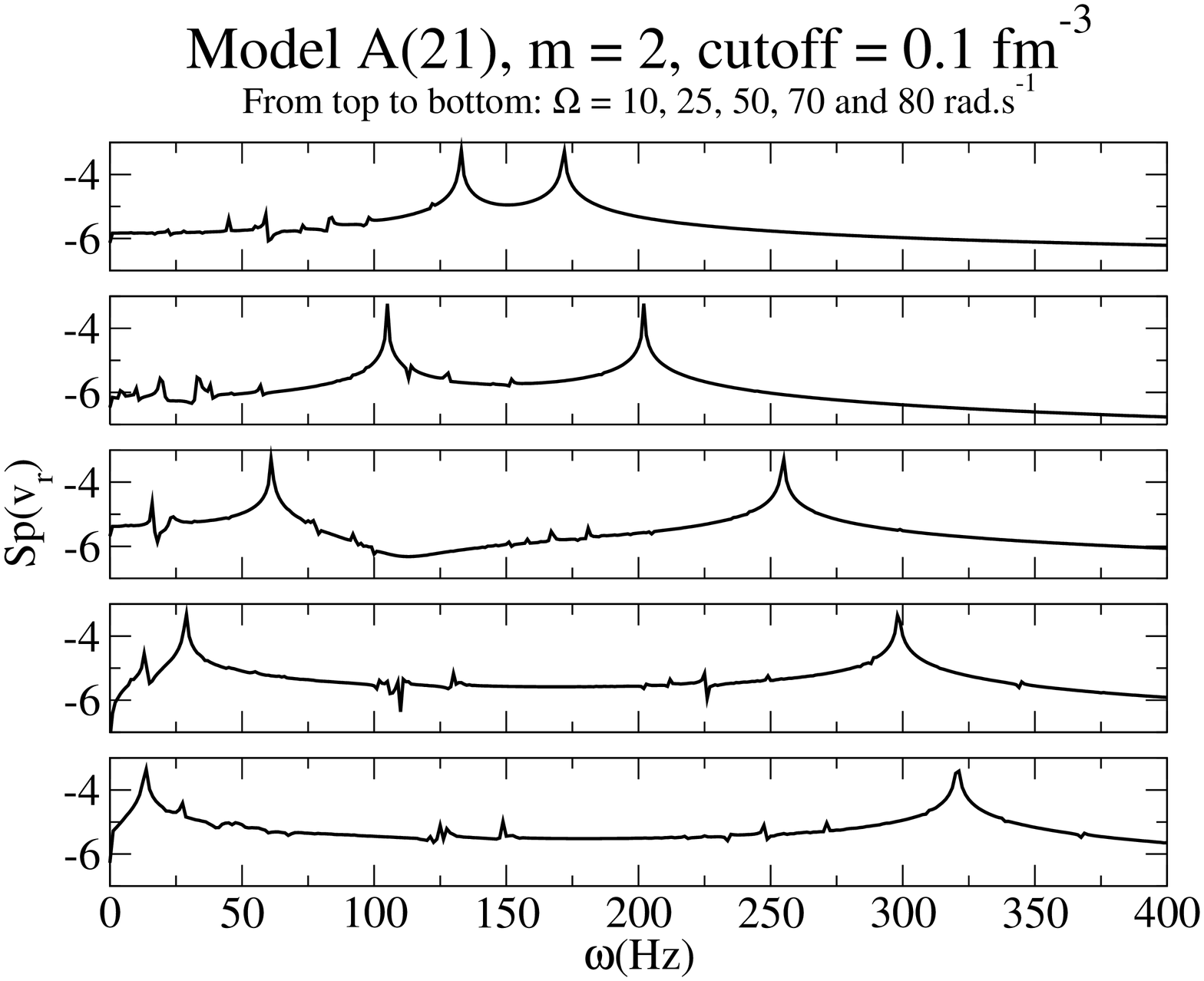}
\caption[]{\label{fig:vr.21.dn.om} Spectra of the radial component of
velocity for several angular velocities. The star is nonbarotropic
and of model A(21), while the initial data are a \mbox{$m\,=\,2$}
density perturbation. See also Figs. \ref{fig:dn.21.dn.om},
\ref{fig:vth.21.dn.om} and \ref{fig:sij.21.dn.om}.}
\end{center}
\end{figure*}

  Fig.\ref{fig:dn.21.dn.om} depicts the spectra of the Lagrangian
perturbation of density. The initial data consist of the same
$m\,=\,2$ perturbation of density [see Eq.(\ref{eq:denspert2})]
without any velocity as in Section \ref{sec:gmod2}. As expected, the
splitting of the {\it g-}modes can be seen, but in addition, we can
observe the emergence of some probable inertial modes: see the peaks
around \mbox{$120$ Hz} and \mbox{$150$ Hz} for \mbox{$\omg\,=\,80$
rad.s$^{-1}$}. The main reasons why these modes can be thought to be
inertial modes are that both their frequencies and their relative
importance in the spectra grow with the angular velocity. Yet, to
better describe them, it is worth to have a look at the spectra of the
velocity in Figs.\ref{fig:vth.21.dn.om} and \ref{fig:vr.21.dn.om},
where we show spectra of the $\th$ and radial components of the
velocity, respectively.\\

 First of all, Fig.\ref{fig:vth.21.dn.om} directly shows that not all
modes were visible in the spectra of the density. Thus, in addition to
the splitted gravity modes and to the already mentioned candidates
inertial modes, there are many others that do not correspond to peaks
in the density spectra. We shall now see how we are lead to classify
them as axial-led inertial modes, following the classification of
Lockitch {\it et al.} \cite{lockitch01}. Obviously, they are inertial
modes since their frequencies increase with the angular velocity. But
our view of classifying them as axial-led is supported by
Fig.\ref{fig:vr.21.dn.om}, which shows that the radial velocity for
those modes is very small comparatively to the radial velocity of
others. Moreover, one of them even seems not to have any radial
counterpart (at this scale of precision): the mode with a frequency of
\mbox{$116$ Hz} for \mbox{$\omg\,=\,80$ rad.s$^{-1}$}, that is to say the mode with
the lowest frequency among those axial-led inertial modes.\\

\begin{figure*}
\begin{center}
\includegraphics[height=15.cm,angle=-90]{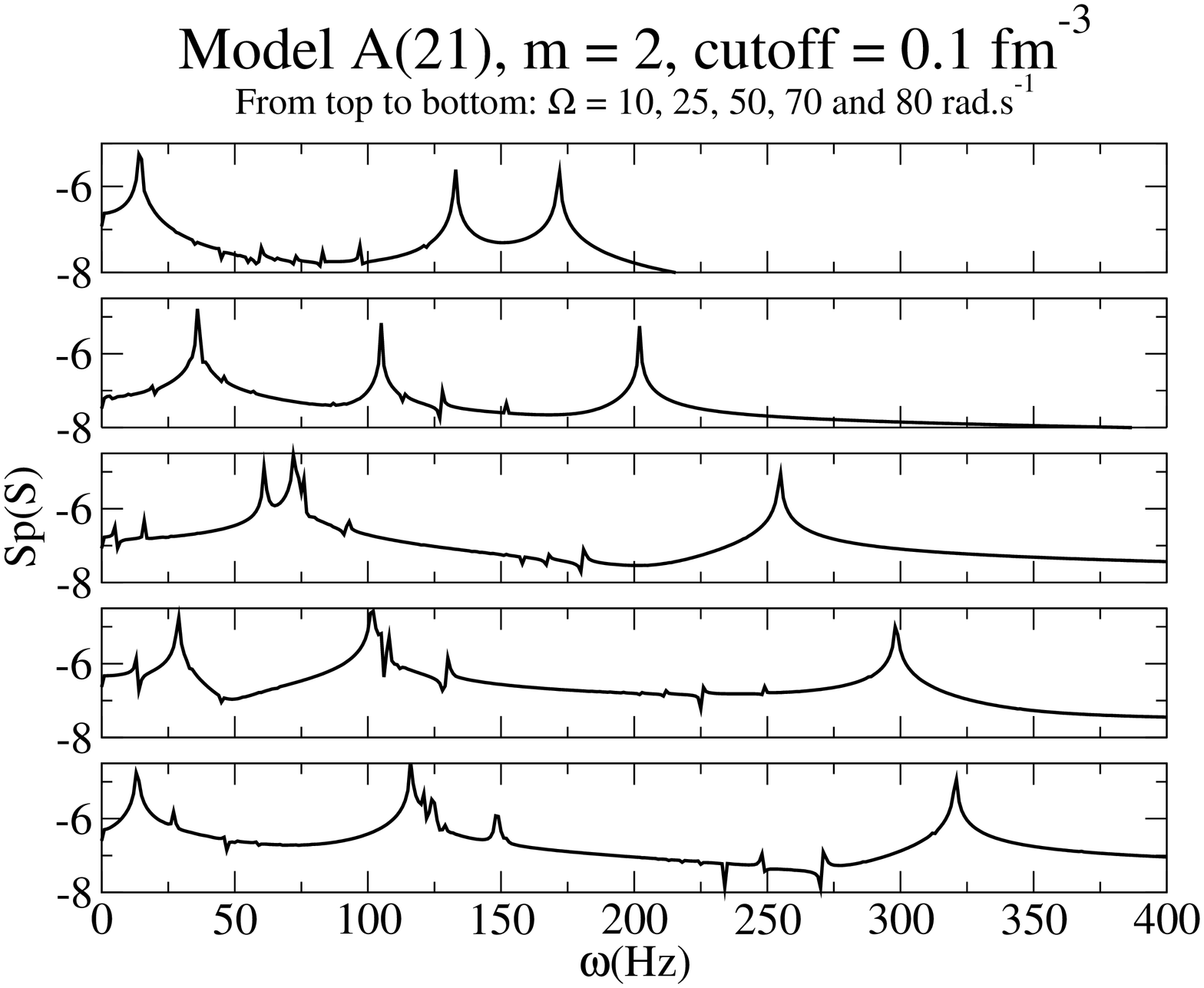}
\caption[]{\label{fig:sij.21.dn.om} Spectra of the current quadrupole
tensor $S$ for several angular velocities. The star is nonbarotropic and
of model A(21), while the initial data are a \mbox{$m\,=\,2$} density
perturbation. See also Figs. \ref{fig:dn.21.dn.om},
\ref{fig:vth.21.dn.om} and \ref{fig:vr.21.dn.om}.}
\end{center}
\end{figure*}

  Interestingly, Fig.\ref{fig:sij.21.dn.om}, which depicts the spectra
  for the current quadrupole tensor, shows that this mode is the mode
  with the largest contribution to this tensor. Notice anyway that its
  amplitude is very low (logarithmic scale) due to the initial
  data. Nevertheless, for all the preceding reasons, we can now say
  that this is the $l\,=\,m\,=\,2$ inertial mode, and that Table
  \ref{tab:bornes} tells us that its frequency is inside the range of
  the continuous spectrum. Indeed, its dimensionless frequency in the
  inertial frame is $1.45$ whereas for model A(21) the range is
  $[1.42,1.6]$. Moreover, with the already displayed spectra of
  velocity, the spectra of a global quantity (the current quadrupole),
  plus various spectra of velocity calculated in several positions
  within the star\footnote{We do not show the corresponding graphs
  here, but they are very similar to what was done in
  Fig.\ref{fig:specrmcutvt} for the barotropic case.}, we verified
  that this is a discrete eigenvalue.\\

  Finally, before we try to investigate the $l\,=\,m\,=\,2$ inertial
  mode in more detail by changing the initial data, let us introduce
  some axial and polar ``kinetic energies'' that shall help us to
  study the coupling between polar and axial parts of the modes. What
  we shall call ``kinetic energy'' in the following was already
  introduced in Paper I and is defined as
\be
\hat{E}\,\,\df\,\frac{1}{2}\,\int_{Star} \!\!\d^3 V \,
  \tilde{n} \,\overrightarrow{W}^2\,,
\ee
where $\tilde{n}\,\df\,n_b\,{N[r]}^2\,a[r]$ [see
Eq.(\ref{e:lin_bar_cons})], $\overrightarrow{W}$ is the three
dimensional vector field (the Eulerian perturbation of velocity)
defined in Eq.(\ref{e:4v_tot}) and $\d^3 V$ is the flat elementary
volume.\\

 With the anelastic approximation, this quantity can be shown to be
 conserved for a barotropic EOS and a vanishing density at the
 surface, conditions that were verified in Paper I. But for
 nonbarotropic EOSs with a cutoff density, two ways to violate this
 conservation exist. The first one, linked to the EOS, corresponds to
 the existence of a ``chemical'' part of the physical energy, while
 the second is linked to fluxes through the surface. To calibrate the
 violation of the conservation due to fluxes, we did time evolutions
 of this kinetic energy in barotropic stars with cutoff densities
 $0.053$, $0.08$ or $0.1$ baryon.fm$^{-3}$. We found that for these
 values, the higher was the cutoff, the better was the conservation,
 which means that our usual cutoff at $0.1$ baryon.fm$^{-3}$ is the
 best of them. Notice that in all cases we saw temporary violations of
 the conservation, but the temporal mean value of the energy was
 always conserved. Moreover, even in the worst cases, the violation
 were never higher than $0.6\%$. On the other hand, testing the
 influence of the ``chemical energy'', we verified that for
 nonbarotropic EOSs with a cutoff at $0.1$ baryon.fm$^{-3}$, the mean
 value was no longer conserved, but the violation was only of the
 order of some $3$ percents. For all these reasons, we shall work in
 the following with this (kinetic) energy as a not too bad indicator
 of the energetic behavior of the oscillations, at least as far as the
 coupling between polar and axial parts is concerned.\\

\begin{figure*}
\begin{center}
\includegraphics[height=15.cm,angle=-90]{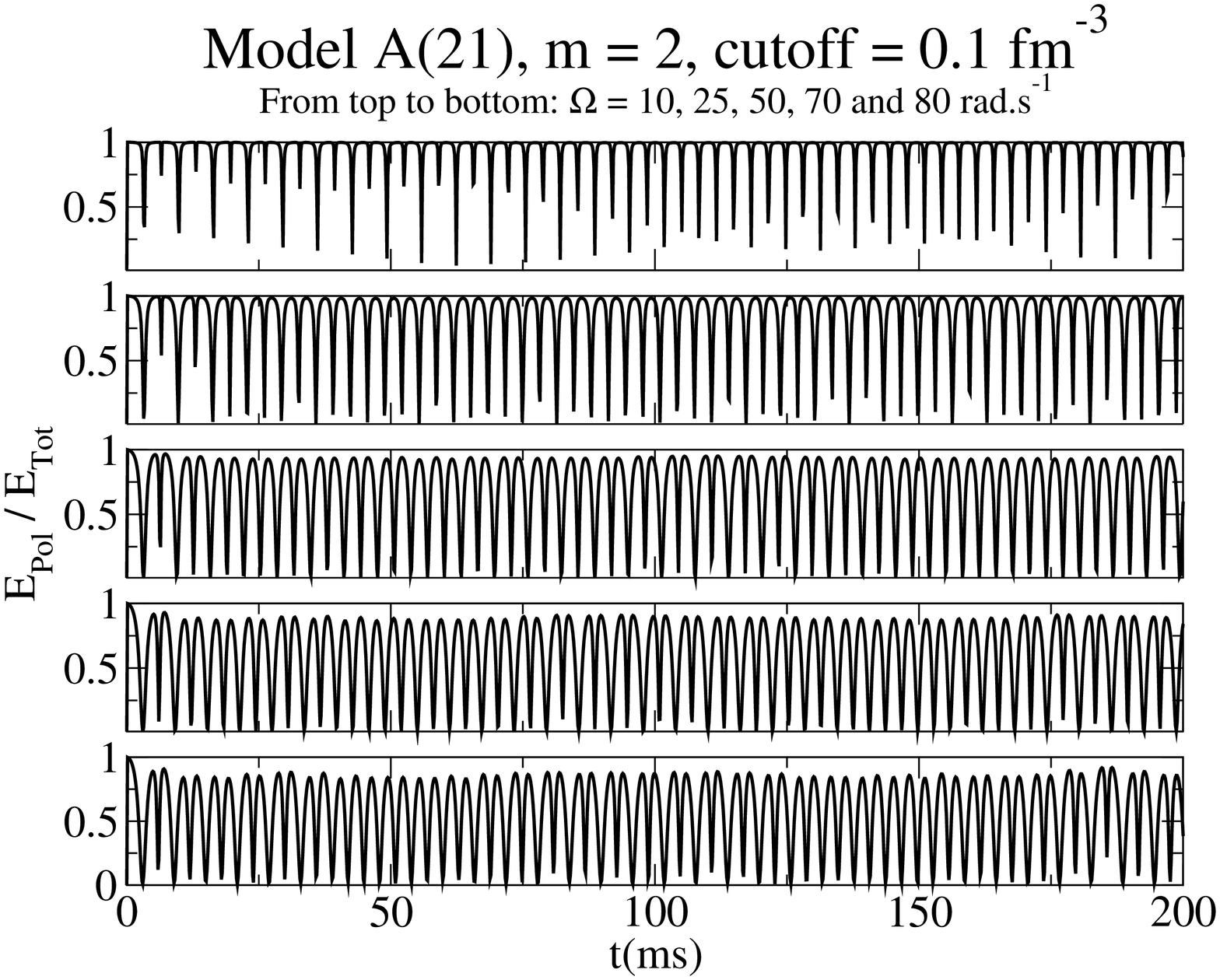}
\caption[]{\label{fig:en.21.dn.om} Time evolution of the ratio between
the polar kinetic energy and the total kinetic energy (as measured in
the rotating frame) in the same star as in previous figures: a
nonbarotropic star of model A(21).}
\end{center}
\end{figure*}

  Indeed, we remind that our algorithm for the solution of
Euler's equations (see Paper I) relies on the use of the radial
velocity and two scalar potentials whose angular divergence and curl
enable us to recover the usual spherical components. By using these
variables, the separation of the polar and axial parts of the velocity
field is done very easily, which naturally leads to the polar and
axial kinetic energies. Thus, Fig.\ref{fig:en.21.dn.om} that displays
the time evolution of the ratio between the polar kinetic energy and
the total kinetic energy give a first insight on the energy flows
between the polar and axial parts of the modes.\\

  In this figure, time evolutions of the ratio are shown, whose
  duration is only $200$ ms in order to make visible the
  ``quasistationnarity'' that is reached after a while. As can be
  seen, starting with a purely polar perturbation, the proportion of
  the energy stored in polar modes is indeed decreasing down to a
  fraction that is between $80$ and $90\,\%$, depending on the angular
  velocity of the star: the faster it rotates, the more energy flows
  to axial modes. Notice anyway that since the initial data are a
  perturbation of density without any velocity, at the very beginning,
  there is, rigorously speaking, no kinetic energy. But in fact, from
  the very first time steps the perturbation of density implies a
  velocity which is initially purely polar. Moreover, we encountered
  quite similar results when we did some tests with as initial data a
  purely polar velocity field without any density perturbation. Only
  the ratio was slightly changed, whilst the time scales were
  identical. Nevertheless, the relevant issue concerning energy flows
  will be how barotropicity affects this evolution. This will be one
  of the topics of the next section.

\subsubsection{Time evolution of the $l\,=\,m\,=2$ inertial mode}

\begin{figure*}
\begin{center}
\includegraphics[height=15.cm,angle=-90]{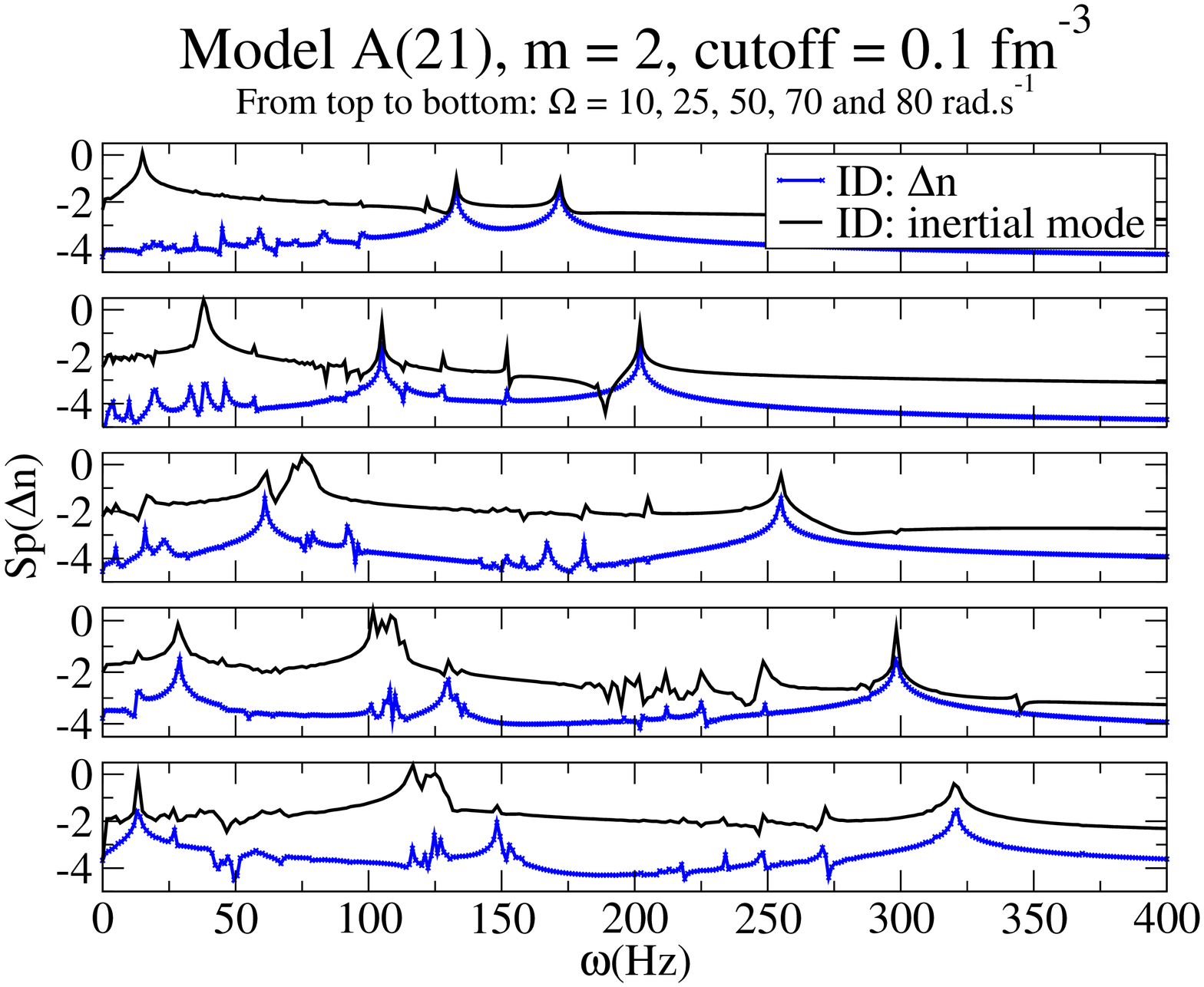}
\caption[]{\label{fig:dn.21.rm.om} Spectra of the density
perturbation for several angular velocities of the star for an initial
\mbox{$l\,=\,m\,=\,2$} inertial mode evolving in a nonbarotropic star
of model A(21).}
\end{center}
\end{figure*}

\begin{figure*}
\begin{center}
\includegraphics[height=15.cm,angle=-90]{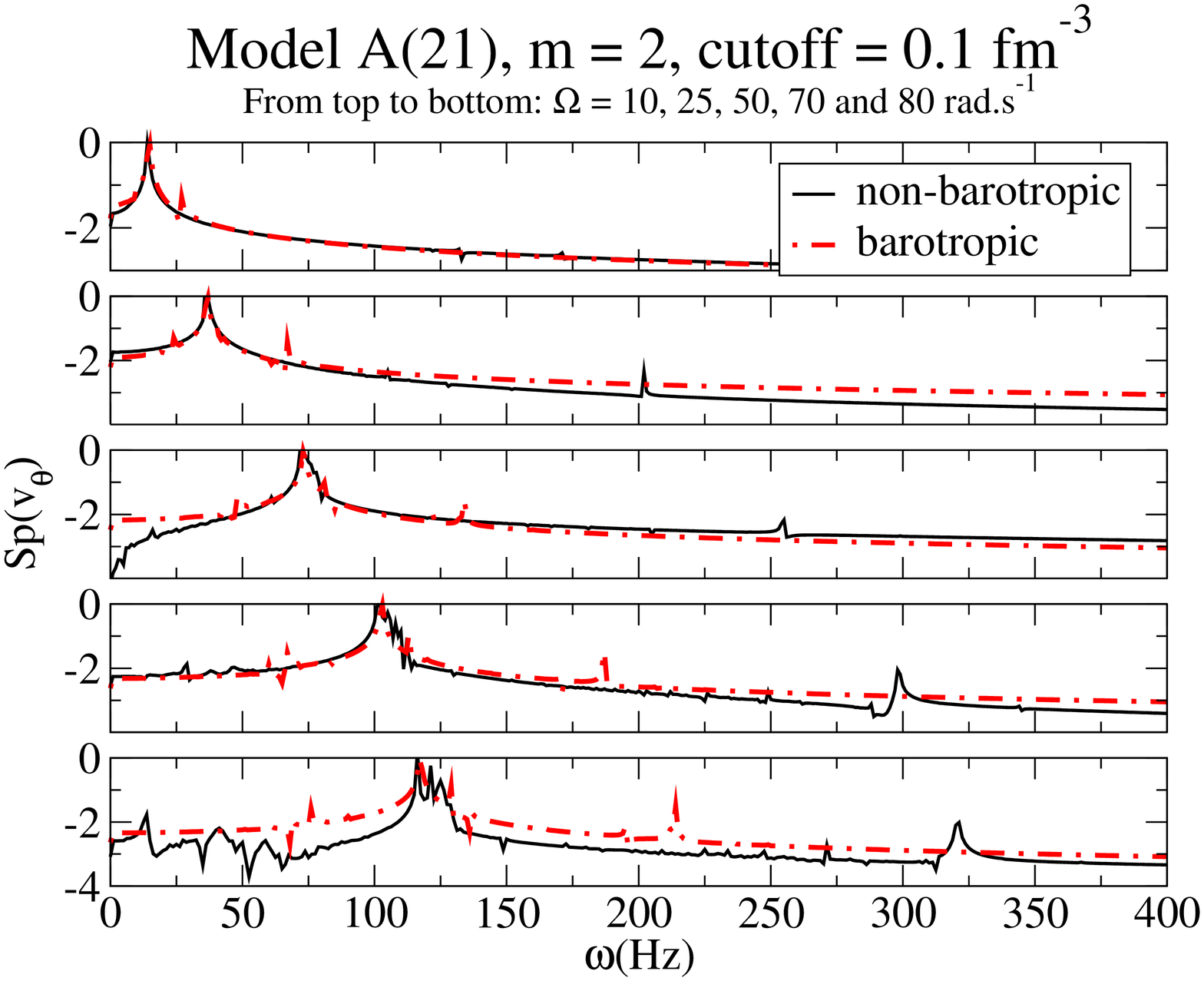}
\caption[]{\label{fig:vth.21.rm.om} Spectra of the $\th$ component of
velocity for several angular velocities in a barotropic and
nonbarotropic star of model A(21) with inertial mode for initial
data.}
\end{center}
\end{figure*}

\begin{figure*}
\begin{center}
\includegraphics[height=15.cm,angle=-90]{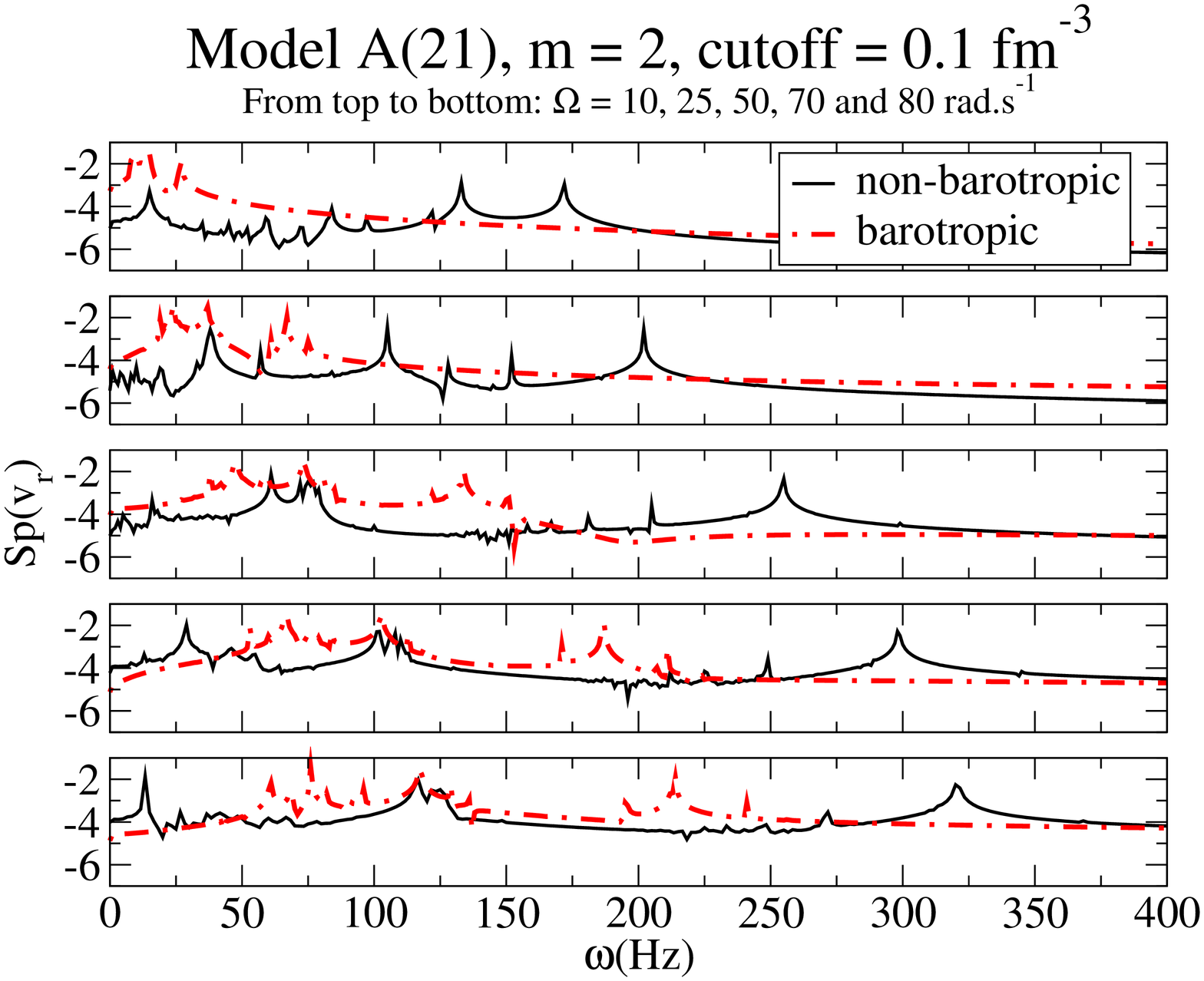}
\caption[]{\label{fig:vr.21.rm.om} Spectra of the radial component of
velocity for several angular velocities in a barotropic and
nonbarotropic star of model A(21) with inertial mode for initial
data.}
\end{center}
\end{figure*}

 We shall now study the time evolution of oscillating
 out-of-equilibrium NSs with different initial data: the $l\,=\,m\,=2$
 inertial mode instead of a density perturbation. In this way, the
 differences with the barotropic case shall be easier to notice, while
 we should also be able to see how much our previous results for
 nonbarotropic stars depend on the initial data. To start with,
 Fig.\ref{fig:dn.21.rm.om} depicts the spectra of the Lagrangian
 perturbation of density for this case, with the results already
 presented in Fig.\ref{fig:dn.21.dn.om} for comparison. The first
 thing to comment on is that even for low angular velocity, gravity
 modes appear, with, as expected, an amplitude that grows with
 $\omg$. But it shall also be pointed out that the $l\,=\,2$ and
 $l\,=\,4$ inertial modes are visible, even if we consider the density
 perturbation spectra, and that their amplitudes stay the same for all
 values of $\omg$. But the coupling between axial and polar parts of
 the velocity field should be better investigated with the spectra of
 the velocity components. These are displayed in
 Figs.\ref{fig:vth.21.rm.om} and \ref{fig:vr.21.rm.om}.\\

 Fig.\ref{fig:vth.21.rm.om} shows that the $\th$ component of the
velocity is hardly affected by the fact that the star is a
barotrop. The main difference between the spectra is indeed only that,
for nonbarotropic stars, gravity modes appear, with amplitudes that
become higher for faster rotating stars. Even the frequency of the
$l\,=\,m\,=\,2$ inertial mode is almost exactly the same. But on the
other hand, the radial part of the velocity
(Fig.\ref{fig:vr.21.rm.om}) is completely changed and quite strongly
depends on the angular velocity, which was predictable. Thus, for
slowly rotating stars, gravity modes dominate and the radial velocity
of the inertial mode is some two orders of magnitude lower than in the
case of a barotropic star. Yet, for increasing angular velocity, the
amplitude and the frequency of the radial part of the $l\,=\,m\,=\,2$
inertial mode become more and more similar to what they are in a
barotropic star. For \mbox{$\omg\,=\,80$ rad.s$^{-1}$}, there is less
than one order of magnitude difference between the barotropic and the
nonbarotropic cases for the $l\,=\,2$ mode (whose frequency in
the inertial frame is \mbox{$w_i\,\sim\,116$ Hz} in this case),
whereas there was two orders for \mbox{$\omg\,=\,10$ rad.s$^{-1}$}. On
the other hand, it should be pointed out that also the amplitude of
the {\it g-}modes grows with $\omg$ (probably meaning that their
``inertial'' part is on the way to dominate buoyancy) while many of
the inertial modes that existed for barotropic stars have
disappeared.\\

\begin{figure*}
\begin{center}
\includegraphics[height=15.cm,angle=-90]{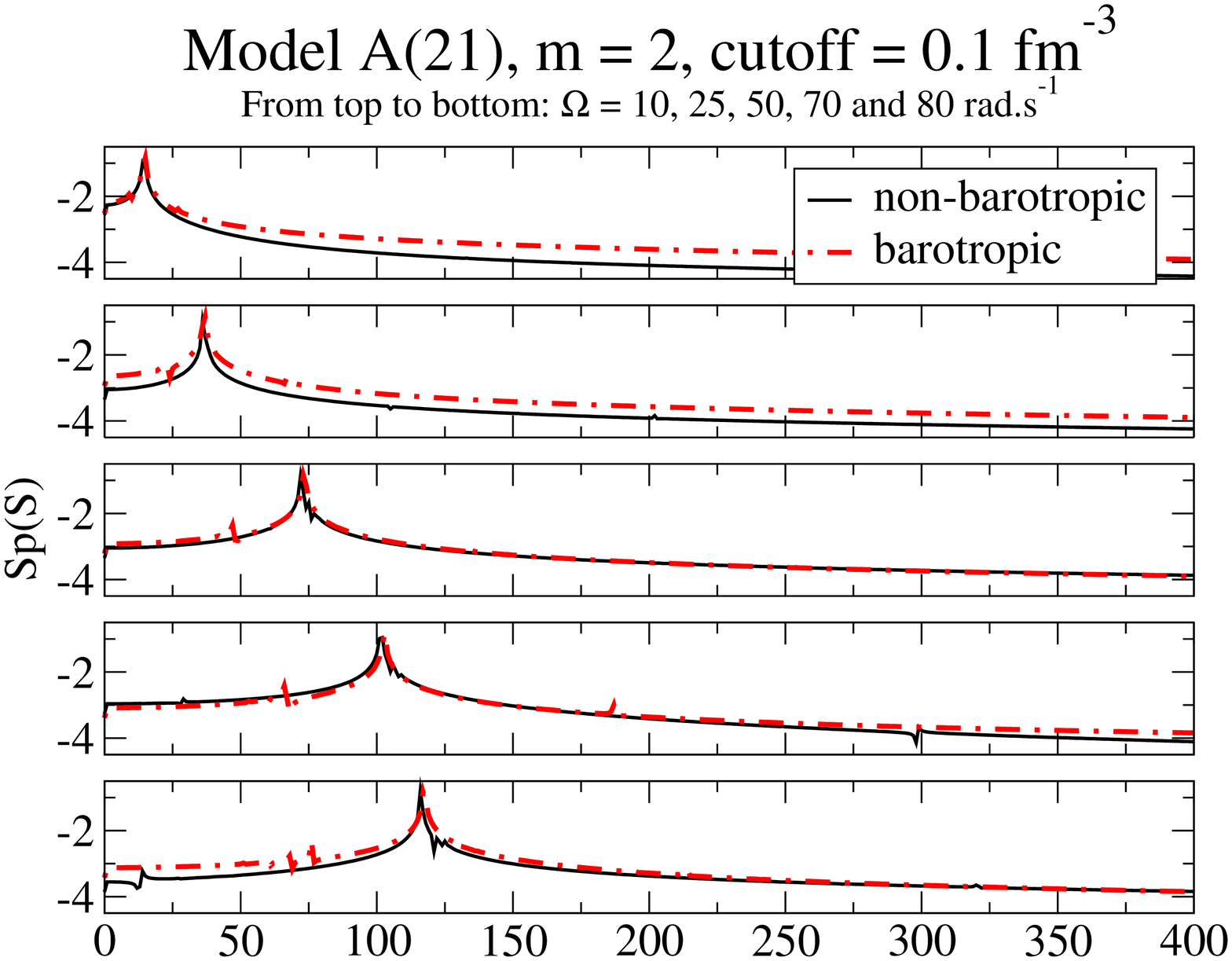}
\caption[]{\label{fig:sij.21.rm.om} Spectra of the current quadrupole
tensor for several angular velocities in a barotropic and
nonbarotropic star of model A(21) with inertial mode for initial
data.}
\end{center}
\end{figure*}

  If we now look at Fig.\ref{fig:sij.21.rm.om} that depicts the
  spectra for the current quadrupole tensor, we shall observe that
  from the point of view of GW emission, barotropicity seems not to have a
  major impact, at least in the spectrum. Indeed, the spectra of this
  tensor do not show remarkable differences, with only a small shift in
  the frequency of the probably most unstable mode that can be pointed
  out. We shall now investigate in more detail how this shift in the
  frequency depends on the EOS, after a brief comment on the time
  evolution of the ratio between polar and axial kinetic energies.\\

\begin{figure*}
\begin{center}
\includegraphics[height=15.cm,angle=-90]{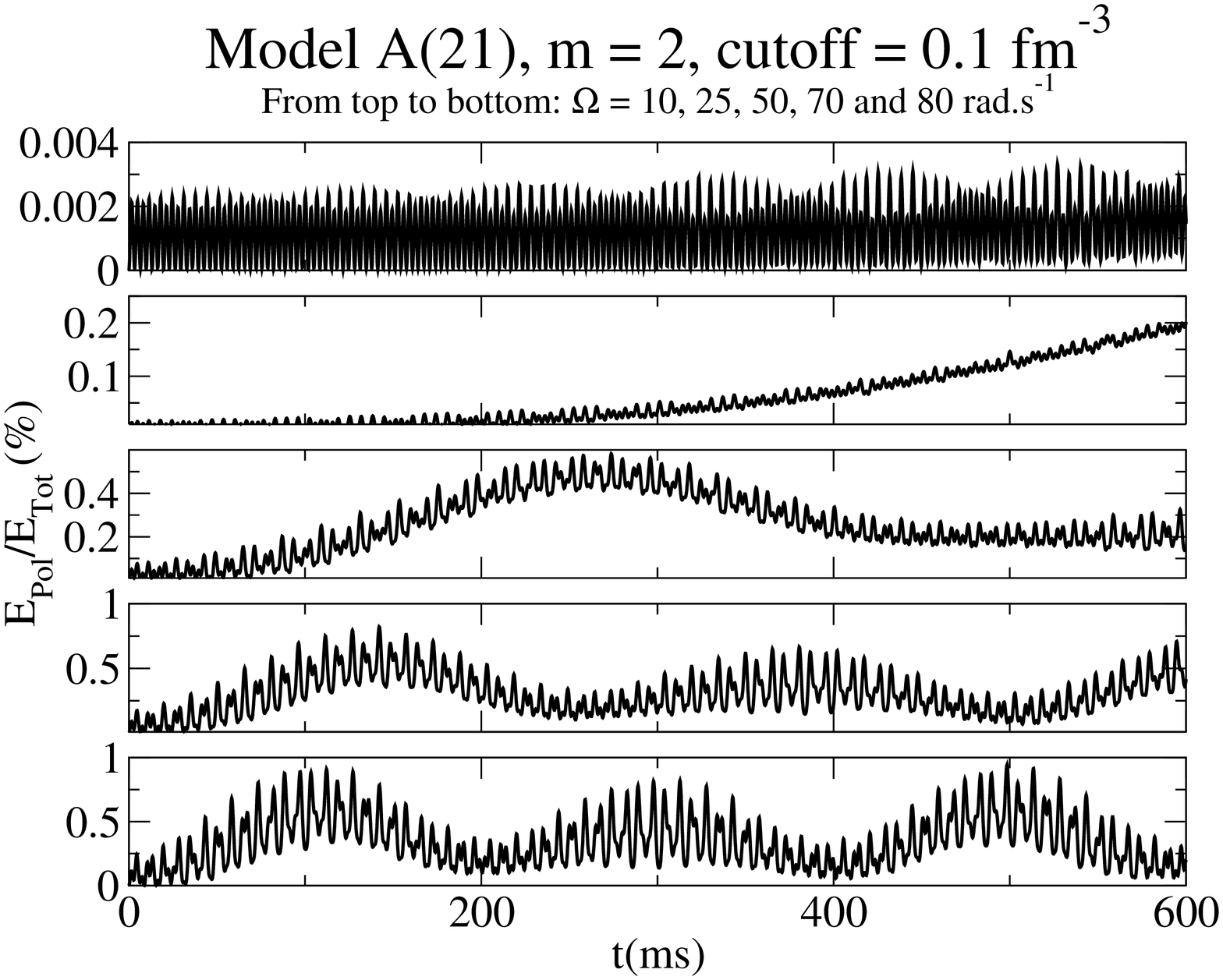}
\caption[]{\label{fig:en.21.rm.om} Time evolution of the ratio between
the polar kinetic energy and the total kinetic energy in the same nonbarotropic star
as in previous figures with inertial mode for initial data.}
\end{center}
\end{figure*}

\begin{figure*}
\begin{center}
\includegraphics[height=15.cm,angle=-90]{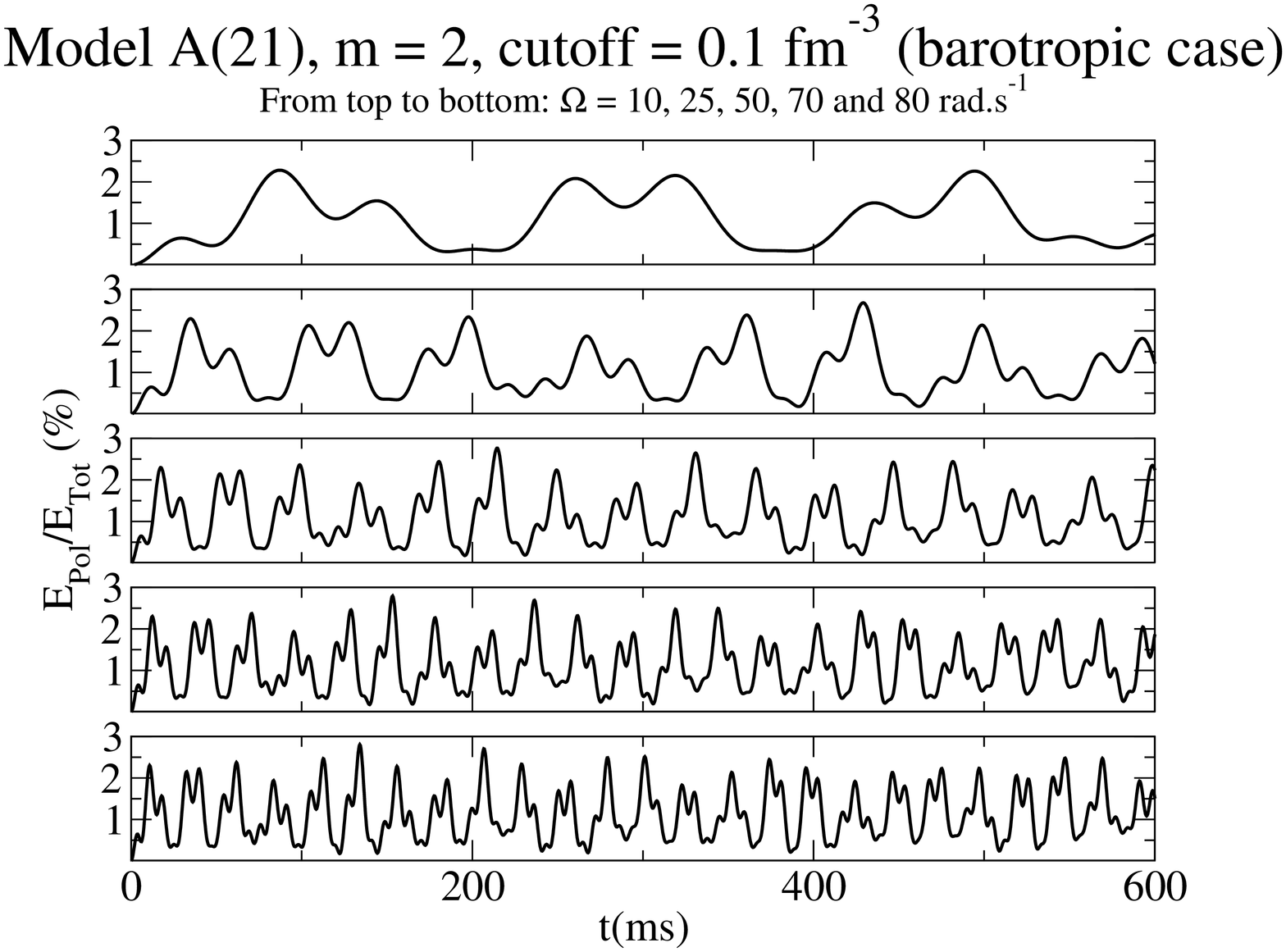}
\caption[]{\label{fig:en.21.baro.om} Time evolution of the ratio between
the polar kinetic energy and the total kinetic energy in the same \underline{barotropic} star
as in previous figures with inertial mode for initial data.}
\end{center}
\end{figure*}

   Indeed, while all previous spectra were very comparable, the time
   evolutions of the polar and axial parts of the kinetic energy are
   quite different for barotropic and nonbarotropic stars. The latter
   situation is illustrated by Fig.\ref{fig:en.21.rm.om}
   (out-of-equilibrium case), and by Fig.\ref{fig:en.21.baro.om}
   (barotropic case)\footnote{Notice that evolutions in the barotropic
   case are, as expected, analogous to the result presented for a
   relativistic polytrop in Paper I.}. Thus, one observes that, for a
   barotropic star, energy is fast flowing back and forth from the
   axial part to the polar part, leading to a kind of ``stationary
   state'' with a medium value of the ratio around $2\,\%$. On the
   other hand, for nonbarotropic stars, the coupling between polar
   and axial parts seems to depend much more on the angular
   velocity. Hence, for $\omg\,=\,10$ or \mbox{$25$ rad.s$^{-1}$},
   even after $600$ ms of evolution, the ``final state'' is still not
   reached and the ratio still grows. But the curves for higher
   angular velocities suggest that they should not get too high since
   the ``stationary situations'' obtained in those cases are
   characterized by ratios lower than for barotropic stars (always
   less than $1\,\%$).\\

   The difference between the typical times is quite easy to
   understand. Indeed, energy flows should have for typical time
   scales the periods of the modes measured in the rotating frame. For
   nonbarotropic stars, whatever the angular velocity, some {\it
   g}-modes exist with periods of some ms, which are then responsible
   for the quite fast flows. However, in the barotropic case, only
   inertial modes can be observed. Since they have periods proportional to the
   inverse of the angular velocity, this is the explanation of the two
   features of the barotropic case: the self-similarity (for different
   values of $\Omega$) of the time evolutions and the acceleration of
   the flows with $\Omega$.\\

   On the other hand, with some caution due to our numerous
   approximations (and also to our definition of ``kinetic energy''),
   we could say that the result concerning amplitude of the ratios can
   be seen as an indication than stratification probably helps the
   unstable inertial mode (the $l\,=\,m\,=\,2$ {\it r-}mode) to be
   driven to instability by GW since its coupling to polar modes
   (which are not driven to instability by the current quadrupole) is
   smaller. Yet, the answer to that question requires a proper
   treatment of the emission of gravitational waves without the
   Cowling approximation, while different viscosities for polar and
   axial modes should also complicate the problem.\\


\begin{figure}
\begin{center}
\includegraphics[height=8.6cm,angle=-90]{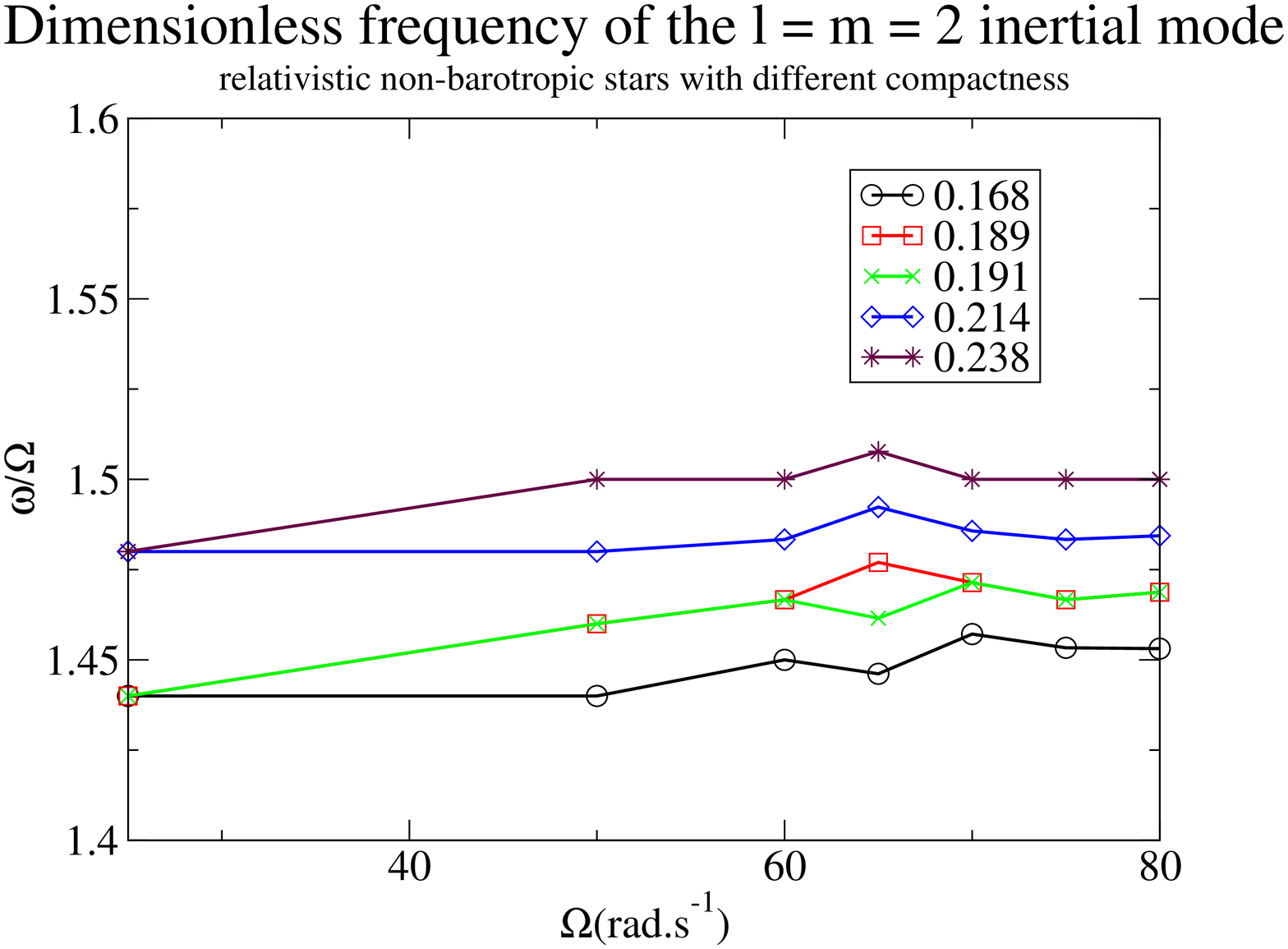}
\caption[]{\label{fig:lm2.nonbaro} Dimensionless frequency versus
angular velocity of the star for \mbox{$l\,=\,m\,=\,2$} inertial mode
in nonbarotropic stars of various compactness. Notice that due to the
fact that evolutions lasted $1$ s, the error bar is equal to
$1\,\rm{rad.s}^{-1}/\omg$. See also Fig.\ref{fig:lm2.baro} for the
barotropic case.}
\end{center}
\end{figure}

\begin{figure*}
\begin{center}
\includegraphics[height=15.cm,angle=-90]{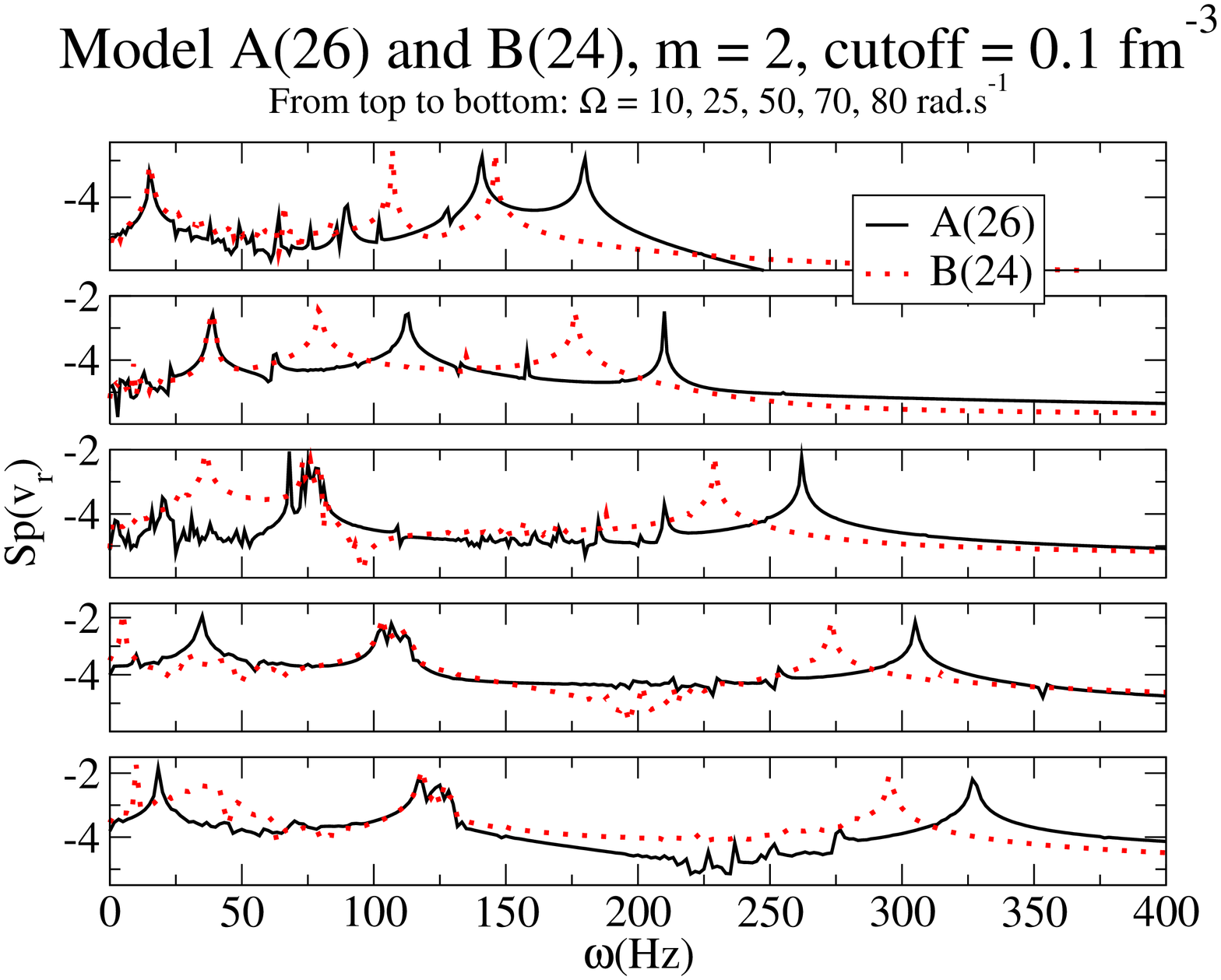}
\caption[]{\label{fig:spec.vr.comp} Spectra of the radial velocity for
several angular velocities in nonbarotropic stars of model A(26) and
B(24) with inertial mode for initial data.}
\end{center}
\end{figure*}

  To conclude this work, we shall give an overview of how the
  frequency of the $l\,=\,m\,=\,2$ inertial mode depends on
  compactness, barotropicity and angular velocity. The latter will be
  the easier: as already discussed, due to the error bars for the
  normalized frequencies, we are not really able to see any
  difference. Moreover, since our study is linear in $\omg$, we expect
  that the normalized frequency of the inertial modes should only
  depend on the compactness of the stars but not on $\omg$. This view
  is supported in the barotropic case and in the nonbarotropic case
  by (respectively) both Figs.\ref{fig:lm2.baro} and
  \ref{fig:lm2.nonbaro}. Thus, the more compact a NS is, the higher
  will be its dimensionless frequency as measured by an inertial
  observer. However, the comparison between these two figures shows
  that taking into account the fact that $npe$ matter should keep a
  frozen composition, which implies a nonbarotropic EOS, leads to
  lower frequencies. Moreover, it is worth pointing out that we
  verified that, for all of our models, the $l\,=\,m\,=\,2$ inertial
  modes had discrete spectra\footnote{which was easier to see by
  looking at the current quadrupole tensor that is a global
  quantity.}, even if Fig.\ref{fig:lm2.nonbaro} and
  Tab. \ref{tab:bornes} demonstrate that all these modes were found
  within the range of the continuous spectra. Nevertheless, as already
  mentioned and as can be seen in Fig.\ref{fig:vr.21.rm.om}, they are
  modes with a radial velocity, which means that they are not purely
  axial. For very slowly rotating stars, this radial component is
  quite small, but it becomes as large as what it is for nonpurely
  axial inertial modes of barotropic stars when the angular velocity
  increases. Finally, notice that even if models A(26) and B(24) have
  the same compactness and then the same frequency for the
  $l\,=\,m\,=\,2$ inertial mode, due to the mixing with gravity modes
  that depends on the detail of the microphysics, their full spectra
  are quite different. This issue is illustrated by
  Fig.\ref{fig:spec.vr.comp} that depicts the spectra of the radial
  component of the velocity for those models with the
  out-of-equilibrium hypothesis. Hence, this probably means that more
  precise studies including more coupling between modes (nonlinear
  terms in $\omg$ or nonlinear corrections to the Euler equations)
  should enable us to make some differences between stars of the same
  compactness, and then to start to probe the inner composition of NSs
  when gravitational waves emitted by their instabilities are
  detected.

\section{Conclusion} \label{sec:conc}

  Inertial modes and gravity modes of slowly rotating relativistic
  stratified neutron stars have been investigated in the linear
  regime. Our study is based on a spectral three dimensional
  evolutionary code which has, at the moment of writing, the restrictions of
  using the Cowling and the anelastic approximations. However, the
  effect of the latter approximation is mainly to kill high frequency
  acoustic modes that should not be strongly coupled to inertial and
  gravity modes for slow rotation. Additionally, the slow rotation
  approximation is reasonable for observed pulsars as, even for the
  fastest known pulsar, the pertinent dimensionless parameter is of
  the order of some percents. The validity of the Cowling
  approximation is more subtle since without perturbations of
  space-time we do not directly include gravitational waves in our
  calculations, whereas they are one of the main motivations for this
  work. Nevertheless, previous studies have proven that the properties
  of inertial modes were not too much affected by its use
  \cite{kokkotas03}.\\

  Another major feature of our work is that we tried to simulate as
  well as possible the microphysical conditions that should occur
  inside not too massive and not too cold neutron stars. Thus, by
  using a quite realistic equation of state for nuclear matter valid
  even when the beta equilibrium is broken, we did the first three
  dimensional time evolution of inertial modes in stratified
  relativistic stars, taking into account the frozen composition of
  every perturbed lump of $npe$ matter. With this approach, we were
  able to show that the coupling between polar and axial modes quite
  strongly increases with the angular velocity. This happens in such a
  way that, in nonbarotropic stars, initially axial inertial modes
  develop a radial part that can be as large as the polar part of the
  general inertial modes of barotropic stars. Yet, several of the
  polar-lead inertial modes of barotropic stars do no longer exist in
  nonbarotropic stars. Moreover, we found that the characteristic
  time scale for the exchanges of energy between polar and axial modes
  is different for barotropic and nonbarotropic stars. Thus, the
  coupling of inertial modes with gravity modes in nonbarotropic
  stars seems to give birth to fast energy interchange between the
  polar and axial parts of the fluid motion. This phenomenon could have
  some implications on the way viscosity or gravitational waves
  emission act, but it would of course need further studies to draw
  robust conclusions.\\

  However, the comparison of inertial modes in stars with and without
  the assumption of frozen composition has already lead us to the
  conclusion that nonbarotropicity decreases the influence of
  compactness on the frequency of the {\it r-}mode, making in addition
  this frequency slightly lower in a nonbarotropic star. If further
  studies with different equations of state and composition support
  this result, it could mean that the measurement of the global
  parameters of a neutron star (compactness, mass and radius if
  absorptions lines are observed, binary system parameters, {\it
  etc.}) in conjunction with the observation of gravitational waves
  emitted by some instabilities of the inertial modes would be very
  instructive about the inner structure of neutron stars. But such a
  Grail would probably need to first improve our understanding of the
  physics with wider explorations of the physical parameters:
  centrifugal terms and deformation of the star, magnetic field,
  differential rotation, nonlinear coupling, {\it etc.} Among them,
  superfluidity is one of the key phenomena, since several studies
  proved that gravity modes are suppressed by its existence. As a
  consequence, even if superfluidity in old neutron stars is already
  strongly supported by both theory and observation (glitches), the
  detailed analysis of the gravitational wave spectra of oscillating
  neutron stars could be the possibility to definitely demonstrate the
  superfluid nature of the neutron stars content by the lack of well
  understood features linked with gravity modes.

\acknowledgments We are indebted with J.A. Pons for his very critical
reading of the manuscript and for his numerous comments, but it is
also a real pleasure to thank N. Andersson and A. Nagar for many useful
suggestions. Finally, we would like to thank the computer department
of the Paris-Meudon Observatory for the technical assistance. This
work has been supported by the EU Programme `Improving the Human
Research Potential and the Socio-Economic Knowledge Base' (Research
Training Network Contract HPRN-CT-2000-00137). PH was partially
supported within the program of European Associated Laboratory
``Astrophysics Poland-France'' (Astro-PF), and LV benefited from the
Jumelage PAN-CNRS Astronomy France-Poland.

\appendix

\section{Dispersion relations} \label{sec:app}

 In this Appendix, the usual basic way to derive the Newtonian
 dispersion relation for {\it g-}modes is summarized, and we briefly discuss
 the effect of the divergence free and anelastic approximations.

\subsection{Usual case with Cowling approximation}

   The easiest way to reach the dispersion relation for gravity
restored modes is based on the perturbative Lagrangian approach. If we
restrict ourselves to linearized equations, and write $\vec{\xi}$ the
displacement vector, $\Dt$ the Lagrangian perturbations and $\dt$ the
Eulerian (linked by the symbolic relation
$\Dt\,\df\,\dt\,+\,\vec{\xi}\ps\vnb$), we have for the Euler equation
\be \label{e:ee}
\pt_t^2\vec{\xi}\,=\,-\frac{1}{n}\,\vnb \dt
P\,-\,\vnb \dt \phi \,-\,\frac{\dt n}{n}\,\vnb \phi\,
\ee
and for the mass conservation equation
\be \label{e:cons_n}
\Dt n\,=\,-\,n\,\vnb\ps\,\vec{\xi}\,,
\ee
where $P$ is the pressure, $\phi$ the gravitational potential and
$\ps$ indicates the usual Euclidean scalar product.\\

  If we now consider only spherical nonrotating stars (only radial
dependence for $\phi$) and neglect the Eulerian perturbation of the
gravitational field (Cowling approximation), the nonradial part of
Eq.(\ref{e:ee}) and Eq.(\ref{e:cons_n}) give us
\be
\Dt n\,=\,n\,\l(\frac{1}{r^2}\,\pt_r\,(r^2\,\xi_r)\,-\,
\frac{l(l\,+\,1)}{r^2\,w^2} \frac{\dt P}{n}\r)\,,
\ee
in which we assumed a decomposition in spherical harmonics of the
scalar variables [$l$ is the index of the spherical harmonic
$Y_l^m(\th,\ph)$] and a harmonic time dependency of $\xi$:
\be
\pt_t^2 \vec{\xi}\,\df\,-\,w^2\,\vec{\xi}\,.
\ee

 Introducing now the equilibrium $c_{eq}$ and frozen composition (adiabatic)
$c_F$ speeds of sound,
\be
|\nb P|\,\df\,\,c_{eq}^2\,|\nb n|\,
\ee
and
\be
\Dt P\,\df\,\,c_F^2\,\Dt n\,,
\ee
with the Brunt-V\"ais\"al\"a frequency defined as
\be
{\cal N}^2\,\df\,\left|\frac{\nb P}{n}\right|^2\,(c_{\beta}^{-2}\,-\,c_F^{-2})\,,
\ee
we obtained the following system of equations
\begin{eqnarray}
\frac{1}{r^2}\,\pt_r\l(r^2\,\xi_r\r)\,+\,\frac{\dt P}{n\,c_F^2}
\l(1\,-\,\frac{w_0}{w}^2\r)\,-\,\frac{g\,\xi_r}{c_F^2}\,=\,0\\
\frac{1}{n}\,\pt_r\dt P\,+\,\frac{g}{c_F^2}\,
\frac{\dt P}{n}\,-\l({\cal N}^2\,-w^2\r)\,\xi_r\,=\,0\,,
\end{eqnarray}
with the Lamb frequency defined as
${w_0}^2\,\df\,c_F^2\,l(l+1)\,/\,r^2$ with $g$ the local gravitational
acceleration.\\

  The new variables
\be
X\,\df\,r^2\,\xi_r\,e^{A}\,,
\ee
\be
Y\,\df\,\dt P\,e^{\,-A}\,,
\ee
with
\be
A\,\df\,-\,\int \frac{g}{c_F^2}\,\d r\,,
\ee
enable to write the system as
\begin{widetext}
\be \label{e:sys}
\pt_r V \, = \left| \begin{array}{cc}
0 & r^2\,e^{2\,A}\,\l(w_0^2\,/\,w^2\,-\,1\r)\,/\,(n\,c_F^2)\\
n\,e^{-2\,A}\,\l(w^2\,-\,{\cal N}^2\r)\,/\,r^2 & 0\\
\end{array} \right|\,V\,,
\ee
\end{widetext}
with
\be \label{e:wkb_v}
V \, \df \l( \begin{array}{ll}
X \\
Y \\
\end{array} \r)\,.
\ee

 A WKB-like approximation,
\be
\pt_r V\,=\,i\,k\,V\,,
\ee
gives the dispersion relation
\be \label{e:disp}
k^2\,=\,\l({\cal N}^2\,-\,w^2\r)\,\frac{1}{c_F^2}\l(\frac{w_0^2}{w^2}\,-\,1\r)\,.
\ee

  Since for neutron stars we have
\be
{\cal N}\,\ll\,\frac{g}{c_F}\,\ll\,c_F\,k\,,
\ee
Eq.(\ref{e:disp}) leads to the two asymptotic and separated branches
for the {\it p-}modes
\be \label{e:disp_p}
w^2\,=\,\frac{c_F}{r}^2\l((k\,r)^2\,+\,l(l+1)\r)\,
\ee
and the {\it g-}modes
\be \label{e:disp_g}
w^2\,=\,{\cal N}^2\,\frac{l(l+1)}{(k\,r)^2\,+\,l(l+1)}\,.
\ee

\subsection{Divergence free and anelastic approximations}

 The two approximations that we shall now discuss differ by the way
they treat the continuity equation (\ref{e:cons_n}). The divergence
free approximation consists, as indicated by its name, in the
replacement of it by the condition
\be
\vnb\ps\vec{\xi}\,=\,0\,,
\ee
while the anelastic approximation neglects the time variation of the
Eulerian perturbation of density, which gives
\be
\vnb\ps \l(n\,\vec{\xi}\r)\,=\,0\,.
\ee

 In the case of the anelastic approximation, some algebra shows that the
system (\ref{e:sys}) is replaced with
\begin{widetext}
\be \label{e:sys_an}
\pt_r V \, = \left| \begin{array}{cc}
0 & r^2\,e^{2\,A}\,w_0^2\,/\,(n\,c_F^2\,w^2)\\
n\,e^{-2\,A}\,\l(w^2\,-\,{\cal N}^2\r)\,/\,r^2 & 0\\
\end{array} \right|\,V\,,
\ee
\end{widetext}
which leads to the relation (\ref{e:disp_g}) as the unique solution:
the {\it p-}modes have been ``killed'' by the anelastic approximation.\\

  On the other hand, it turns out that the same exercise with the
divergence free approximation also gives the relation (\ref{e:disp_g})
as the unique solution, but the vector on which the WKB approximation
is done has to be defined as
\be 
V \, \df \l( \begin{array}{ll}
X \,\df \, r^2\,\xi_r\\
Y \,\df \, \dt P\,e^{-\,A}\\
\end{array} \r)\,.
\ee

  This small difference in the definition of the second component of
the vector between the divergence free approximation and the ``exact
solution'' seems to indicate that the anelastic approximation (for
which there is not such a difference) is closer to the exact solution
than the divergence free approximation is.

%
\def\aa{Astron. {\rm \&} Astrophys. }                          
\def\aapr{Astron.{\rm \&}Astrophys., Rev. }                      
\def\aaps{Astron.{\rm \&}Astrophys., Supp. }                     

\def\apj{Astrophys. J. }                         
\def\apjl{Astrophys. J., Lett. }                 
\def\apjs{Astrophys. J., Supp. }                 

\def\cqg{Class. Quant. Grav. }                   

\def\mnras{Mon. Not. of the Royal Astron. Soc. }         
        
\def\apss{Astron. and Space Sci. }               

\def\jfm{J. of Fluid Mech. }
\def\ijmpd{Int. J. of Mod. Phys. D }
\def\jas{J. Atm. Sci. }
\def\nat{Nature }                                        

\def\nphys{Nucl. Phys. }                                 
\def\nphysa{{Nucl. Phys. A }}                            
\def\nphysb{{Nucl. Phys. B }}                            

\def\ptrsla{Phil. Trans. R. Soc. Lond. A }
\def\phmag{Phil. Mag. }
\def\physrep{Phys. Rep. }                                
\def\physl{Phys. Lett. }                                 
\def\physla{Phys. Lett. A }                              
\def\physlb{Phys. Lett. B }                              

\def\pr{Phys. Rev. }                          
\def\pra{Phys. Rev. A: General Physics }              
\def\prb{Phys. Rev. B: Solid State }          
\def\prc{Phys. Rev. C }                               
\def\prd{Phys. Rev. D }                               
\def\pre{Phys. Rev. E }                               
\def\prl{Phys. Rev. Lett. }                       
\def\qjrms{Quart. J. Roy. Meteor. Soc. }
\def\sicomp{SIAM J. on Comp. }                           
\def\jcomp{J. of Comp. Phys. }             
\def\jcam{J. of Comp. and App. Math. }     
\def\ssr{Space Sci. Rev. }                       

\end{document}